\documentclass[10pt,notitlepage,onecolumn,showpacs,preprintnumbers,amsmath,amssymb,superscriptaddress, pra,floatfix]{revtex4-1}
\usepackage{bm}
\usepackage{graphicx}
\usepackage{dcolumn}
\usepackage{tabularx}
\usepackage{bm}
\usepackage{epstopdf}
\usepackage{amsmath}
\usepackage{color}
\usepackage{here}
\usepackage{setspace}
\usepackage[left=37.5truemm,right=37.5truemm]{geometry}

\begin{document}
\setstretch{1.0}
 
\preprint{Preprint}
\title{Designs toward synchronization of optical limit cycles with coupled silicon photonic crystal microcavities}

\author{N. Takemura} 
\email[E-mail: ]{naotomo.takemura.ws@hco.ntt.co.jp}
\affiliation{Nanophotonics Center, NTT Corp., 3-1, Morinosato Wakamiya Atsugi, Kanagawa 243-0198, Japan}
\affiliation{NTT Basic Research Laboratories, NTT Corp., 3-1, Morinosato Wakamiya Atsugi, Kanagawa 243-0198, Japan}
\author{M. Takiguchi}
\author{M. Notomi}
\affiliation{Nanophotonics Center, NTT Corp., 3-1, Morinosato Wakamiya Atsugi, Kanagawa 243-0198, Japan}
\affiliation{NTT Basic Research Laboratories, NTT Corp., 3-1, Morinosato Wakamiya Atsugi, Kanagawa 243-0198, Japan}


\begin{abstract}
A driven high-Q Si microcavity is known to exhibit limit cycle oscillation originating from carrier-induced and thermo-optic nonlinearities. We propose a novel nanophotonic device to realize synchronized optical limit cycle oscillations with  coupled silicon (Si) photonic crystal (PhC) microcavities. Here, coupled limit cycle oscillators are realized by using coherently coupled Si PhC microcavities. By simulating coupled-mode equations, we theoretically demonstrate mutual synchronization (entrainment) of two limit cycles induced by coherent coupling. Furthermore, we interpret the numerically simulated synchronization in the framework of phase description. Since our proposed design is perfectly compatible with current silicon photonics fabrication processes, the synchronization of optical limit cycle oscillations will be implemented in future silicon photonic circuits. 
\end{abstract}
\maketitle

\section{Introduction}
Synchronization is a universally observed phenomenon in nature \cite{Pikovsky2003}. In fact, the observation of synchronization has a long history, which may go back to the 17th century with Huygens's discovery of synchronization of two pendulum clocks. In the 19th century, Lord Rayleigh reported the unison of sounds in acoustical systems. The first modern experimental studies of synchronization were performed by Appleton and van der Pol in the early 20th century using electrical and radio engineering techniques \cite{Appleton1922,VanDerPol1927}. On the other hand, for a clear understanding of synchronization, we had to wait until the late 20th century, when phase description of limit cycles was developed by Winfree and Kuramoto \cite{Winfree1967,Kuramoto2003}. Limit cycle oscillation emerges from a nonlinear dissipative system and well models various rhythm and self-pulsing phenomena. Since limit cycles have stable orbits, they are different from harmonic oscillations in conservative systems. The main idea of phase description is to describe limit cycle dynamics solely with a (generalized) phase degree of freedom. The phase description was found to be a powerful tool for understanding not only single limit cycle dynamics but also synchronization phenomena. In fact, for an intuitive understanding of mutual synchronization (entrainment) of coupled limit cycles, phase description provides a powerful tool called the phase coupling function. Furthermore, phase description is not limited to two oscillators, and it can also be used to analyze an ensemble of coupled oscillators, which is called the Kuramoto model. Nowadays, the phase analysis of synchronization is an indispensable tool to understand various synchronization phenomena in physics, chemistry, biology, and physiology. In biology, the numerous examples of synchronization phenomena range from the circadian rhythm to firefly synchronization \cite{Pikovsky2003}. In physics, synchronization phenomena in several systems has only recently been discussed. The most famous example may be the Josephson junction array, which is known to be described by the Kuramoto model \cite{Tsang1991,Wiesenfeld1996,Barbara1999}. In photonic systems, synchronization has been demonstrated with coupled lasers, microcavity polaritons, optomechanical oscillators, and trapped ions \cite{Thornburg1997,Hohl1999,Kozyreff2000,Allaria2001,Baas2008,Zhang2012,Bagheri2013,Lee2013,Walter2014,Ohadi2016}. Furthermore, very recently, a frequency comb was interpreted in terms of synchronization \cite{Hillbrand2020}.

In this paper, we propose a novel nanophotonic system with standard silicon (Si) photonic crystal (PhC) technologies that realizes synchronization of optical limit cycles. In our previous paper \cite{Takemura2020}, we experimentally investigated the detailed properties of stochastic limit cycle oscillation (self-pulsing) in a single driven high-Q Si PhC microcavity. Here, we extended the previous study to coupled driven Si PhC microcavities. First, by numerically simulating coupled-mode equations, we demonstrate that introducing coherent field coupling between two cavities gives rise to synchronization (entrainment) of two limit cycle oscillations. Interestingly, we found that the synchronization phase (for example, in- and anti-phase synchronizations) can be controlled by the phase difference between two laser inputs. Second, we qualitatively interpreted the numerically demonstrated synchronization in the framework of the phase description (phase reduction theory). For this purpose, we calculated the  phase coupling function, which plays a central role in phase description \cite{Stankovski2017,Kuramoto2003}. The obtained phase coupling function intuitively explains the origin of the synchronization and the synchronization phase. Finally, we demonstrated synchronization in a realistic coupled cavity device, which has moderately different cavity resonance frequencies. 

PhC cavity structures largely enhances carrier-induced and thermo-optic optical nonlinearities with their very high-$Q$ value and nanoscale mode-volume \cite{Barclay2005,Uesugi2006,Leuthold2010}.
Employing the enhanced carrier-induced and thermo-optic nonlinearities in high-Q PhC cavities, optical bistability \cite{Tanabe2005,Notomi2005,Weidner2007,Haret2009,Rossi2009}, limit cycle oscillation \cite{Cazier2013,Yacomotti2006,Brunstein2012}, and excitability \cite{Yacomotti2006,Brunstein2012} were demonstrated. 
Furthermore, recently, coupled PhC cavities has been actively investigated to realize, for instance, slow-light \cite{Matsuda2011}, Fano resonance \cite{Yang2009,Nozaki2013}, unconventional photon blockade \cite{Flayac2015}, and self-pulsing coupled nanolasers \cite{Yacomotti2013,Yu2017,Marconi2020}. 
Here, Si PhC cavities are advantageous also for studying synchronization of optical limit cycles from the standpoint of measurements and their controllability. 
In particular, for measurements, the real-time  dynamics of light outputs are easily obtained with conventional optical setups. Meanwhile, the input power and frequency of a driving laser are easily controlled. Furthermore, since the proposed coupled Si PhC cavity device does not require any active material, and is based on the standard Si fabrication technique, its integration with  other Si photonic devices is easy. Thus, it will be easy to implement the demonstrated limit cycle synchronization for future silicon photonic information processing and optical communications \cite{Bregni2002}. Ultimately, an array of Si PhC cavities will work as a one-dimensional nearest-neighbor coupled Kuramoto model.
\begin{figure}
\centering\includegraphics[width=11cm]{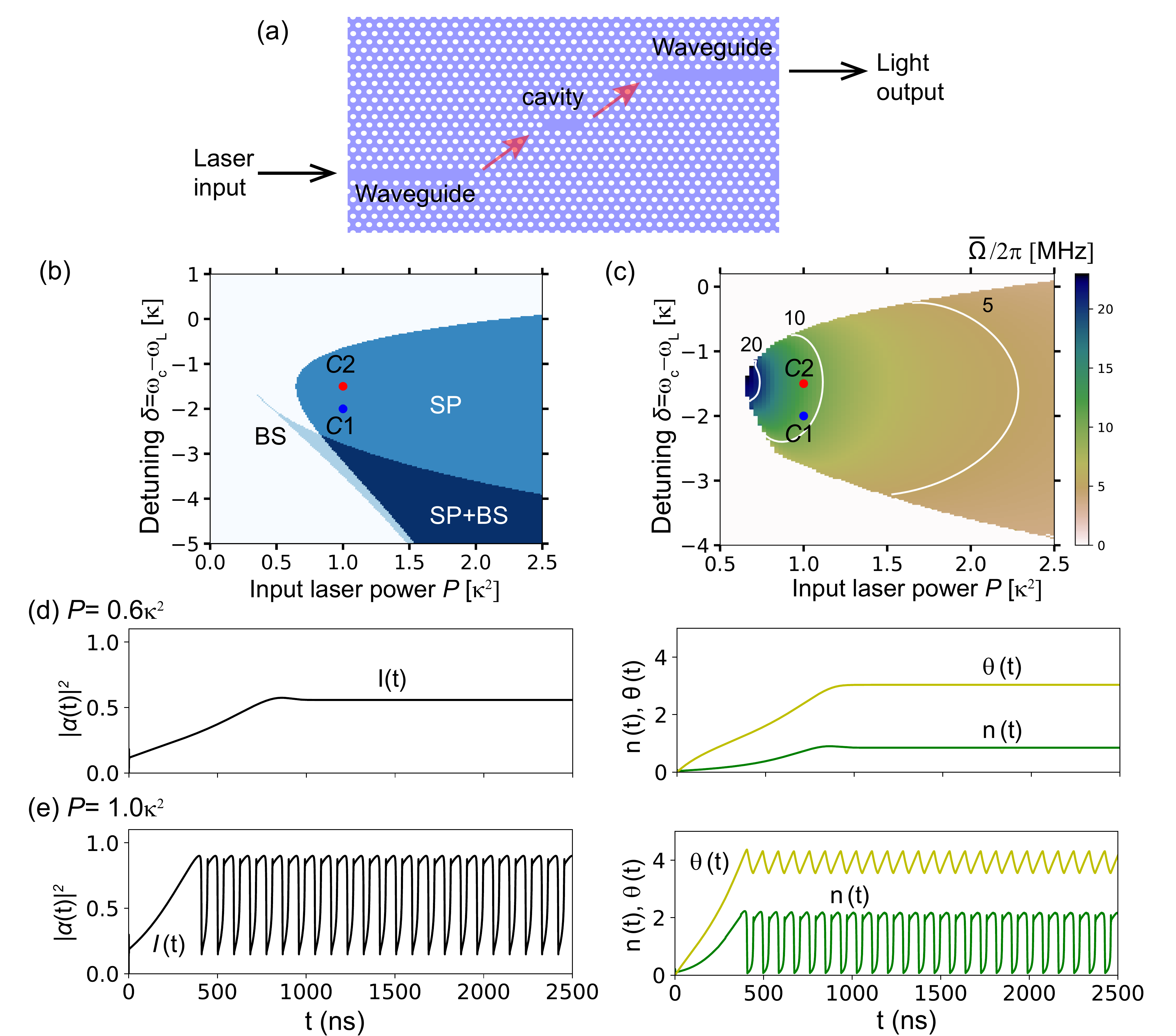}
\caption{(a)  Schematic of a high-Q Si PhC microcavity with two waveguides. (b) Self-pulsing (SP) and bistable (BS) regions as functions of laser input power $P$ and detuning $\delta(=\omega_L-\omega_c)$. (c) Input power $P$ and detuning $\delta$ dependence of limit cycle oscillation frequency $\Omega$. In (b) and (c), blue and red filled circles represent parameters used for the cavity $C1$ and $C2$ in Fig. \ref{fig:schematic_coupled}, respectively. (d), (e) Time evolutions of the light output intensity $|\alpha(t)|^2$ (left), carrier $n(t)$ (right), and thermal component $\theta(t)$ (right) for $P=0.6\kappa^2$ (d) and $1.0\kappa^2$ (e). In (d) and (e), we used $\delta=-2\kappa$.}
\label{fig:schematic_single}
\end{figure} 

\section{Limit cycle in a single high-Q Si PhC cavity}
First, we review limit cycle oscillation emerging from a single high-Q Si PhC cavity, which we investigated in our previous paper \cite{Takemura2020}. We consider a single Si L3-type cavity with two waveguides as schematically shown in Fig. \ref{fig:schematic_single}(a), which is the same as in Ref \cite{Takemura2020}. The PhC slab is a two-dimensional hexagonal lattice, and the cavity is introduced by removing three air-holes. Note that, in the sample used in Ref. \cite{Takemura2020}, several air-holes around the cavity region were carefully modulated to achieve larger $Q$ value than that of the conventional L3 cavity \cite{Kuramochi2014}. The cavity, which has resonance frequency $\omega_c$, is driven by a laser input with frequency $\omega_L$ and power $P$ through the input waveguide. When input power exceeds a critical value, the output light  exhibits limit cycle oscillation (self-pulsing) originating from nonlinear field, carrier, and thermal dynamics. 

Now, we write up  the coupled-mode equations describing field, carrier, and thermal dynamics in the nonlinear Si PhC cavity, which were proposed in Ref. \cite{VanVaerenbergh2012,Zhang2013} and also used in our previous paper \cite{Takemura2020}. Electric field $\alpha$, normalized carrier density $n$, and normalized thermal effect $\theta$ follow the coupled-mode equations 
\begin{eqnarray}
\dot{\alpha}&=&\kappa\lbrace i(-\delta/\kappa-\theta+n)-(1+fn)\rbrace\alpha+\sqrt{P}\label{eq:coupled_alpha}\\
\dot{n}&=&-\gamma n+\kappa\xi|\alpha|^4\label{eq:coupled_n}\\
\dot{\theta}&=&-\Gamma\theta+\kappa\beta|\alpha|^2+\kappa\eta|\alpha|^2n,\label{eq:coupled_theta}
\end{eqnarray}
where the detuning $\delta$ is defined as $\delta=\omega_L-\omega_c$. The thermal effect $\theta$ is proportional to a temperature difference between the internal and external  regions of the cavity. It is important to note that the variables $n$ and $\theta$ are normalized so that the nonlinear coefficients before $n$ and $\theta$ in Eq. (\ref{eq:coupled_alpha}) are unity. The nonlinear coefficients $f$, $\xi$, $\beta$, and $\eta$ represent free-carrier absorption (FCA), two-photon absorption (TPA), heating with linear photon absorption, and FCA-induced heating, respectively. The small Kerr nonlinearity is neglected in the coupled-mode equations. In the rest of this paper, we use $f=0.0244$, $\xi=8.2$, $\beta=0.0296$, and $\eta=0.0036$, which are the same as in Ref \cite{Zhang2013}. Although a precise determination of the values of the nonlinear coefficients is difficult, exact values are not necessary, and the qualitative reproduction of the observed limit cycle oscillation is sufficient. For the lifetimes of the three variables, we set $1/2\kappa=300$ ps ($Q\sim3.5\times10^5$), $1/\gamma=200$ ps, and $1/\Gamma=100$ ns. As discussed in Ref. \cite{Tanabe2005,Tanabe2008}, in the L3-type PhC cavity, due to the small cavity region, fast carrier diffusion makes the carrier lifetime comparable to the field lifetime. The details of our model are described in the Supplemental Material in Ref. \cite{Takemura2020}. 

Here, we briefly discuss the steady-state properties of coupled-mode equations (\ref{eq:coupled_alpha})-(\ref{eq:coupled_theta}). Here, $\alpha_{\rm ss}$, $n_{\rm ss}$, and $\theta_{\rm ss}$ represent the steady state values of the field, carrier, and thermal effect, respectively. By setting $\dot{\alpha}=0$, $\dot{n}=0$, and $\dot{\theta}=0$ in Eqs (\ref{eq:coupled_alpha})-(\ref{eq:coupled_theta}), an algebraic equation for $I_{\rm ss}=|\alpha_{\rm ss}|^2$ is obtained as
\begin{eqnarray}
0=f_{\rm ss}(I)\equiv I\left[\left(-\frac{\delta}{\kappa}-\frac{\kappa}{\Gamma}\beta I-\frac{\kappa^2}{\gamma\Gamma}\eta\xi I^3+\frac{\kappa}{\gamma}\xi I^2\right)^2+\left(1+\frac{\kappa}{\gamma}f\xi I^2\right)^2\right]-\frac{P}{\kappa^2}.\label{eq:cubic}
\end{eqnarray}
Depending on input power $P$ and detuning $\delta$, the algebraic equation (\ref{eq:cubic}) has one or two solutions for $I$. We numerically solve Eq. (\ref{eq:cubic}) and obtain $I_{\rm ss}$. Using $I_{\rm ss}$, we respectively calculate $n_{\rm ss}$ and $\theta_{\rm ss}$ as 
\begin{eqnarray}
n_{\rm ss}=\frac{\kappa}{\gamma}\xi I_{\rm ss}^2\ \ \ {\rm and}\ \ \theta_{\rm ss}=\frac{\kappa}{\Gamma}\beta I_{\rm ss}+\frac{\kappa^2}{\gamma\Gamma}\eta\xi I_{\rm ss}^3.
\end{eqnarray}
Using $n_{\rm ss}$ and $\theta_{\rm ss}$, we can write the complex electric field $\alpha_{\rm ss}$ as
\begin{eqnarray}
\alpha_{\rm ss}=\frac{\sqrt{P}}{\kappa}\cdot\frac{(1+fn_{\rm ss})+i(-\delta/\kappa-\theta_{\rm ss}+n_{\rm ss})}{(-\delta/\kappa-\theta_{\rm ss}+n_{\rm ss})^2+(1+fn_{\rm ss})^2}.
\label{eq:ss_alpha}
\end{eqnarray}
Second, to check the stabilities of the steady states, we perform a linear stability analysis for coupled-mode equations (\ref{eq:coupled_alpha})-(\ref{eq:coupled_theta}). For this purpose, decomposing the complex field $\alpha$ as $\alpha=x+iy$, we rewrite Eqs. (\ref{eq:coupled_alpha})-(\ref{eq:coupled_theta}) as
\begin{eqnarray}
\dot{\bm x}=
{\bm f}({\bm x})=
\left( \begin{array}{c}
f_x({\bm x})\\
f_y({\bm x})\\
f_n({\bm x})\\
f_\theta({\bm x})
\end{array} \right)
=
\left( \begin{array}{c}
 -\kappa(1+f n) x-\kappa(-\delta/\kappa-\theta+n) y+\sqrt{P}\\\
 -\kappa(1+f n) y+\kappa(-\delta/\kappa-\theta+n) x\\
-\gamma n+\kappa\xi (x^2+y^2)^2\\
-\Gamma\theta+\kappa\beta (x^2+y^2)+\kappa\eta (x^2+y^2)n
\end{array} \right),
\label{eq:stanard_f}
\end{eqnarray}
where the vector ${\bm x}$ is defined as ${\bm x}=(x,y,n,\theta)$. Now, a 4$\times$4 Jacobian matrix corresponding to the dynamical system in Eq. (\ref{eq:stanard_f}) is given by \renewcommand{\arraystretch}{1.2}
\begin{eqnarray}
{\bm J}({\bm x})
=
\left( \begin{array}{cccc}
\frac{\partial f_x}{\partial x} & \frac{\partial f_x}{\partial y} & \frac{\partial f_x}{\partial n} & \frac{\partial f_x}{\partial \theta}\\
\frac{\partial f_y}{\partial x} & \frac{\partial f_y}{\partial y} & \frac{\partial f_y}{\partial n} & \frac{\partial f_y}{\partial \theta}\\
\frac{\partial f_n}{\partial x} & \frac{\partial f_n}{\partial y} & \frac{\partial f_n}{\partial n} & \frac{\partial f_n}{\partial \theta}\\
\frac{\partial f_\theta}{\partial x} & \frac{\partial f_\theta}{\partial y} & \frac{\partial f_\theta}{\partial n} & \frac{\partial f_\theta}{\partial \theta}
\end{array} \right)
=\kappa
\left( \begin{array}{cccc}
-f n - 1 & \delta/\kappa - n + \theta & -f x - y & y\\
-\delta/\kappa + n - \theta & -f n - 1 & -f y + x & -x\\
4\xi x(x^2 + y^2) & 4\xi y (x^2 + y^2) & -\gamma/\kappa & 0\\
2 \beta x + 2 \eta n x & 2 \beta y + 2 \eta n y & \eta (x^2 + y^2) & -\Gamma/\kappa
\end{array} \right).\nonumber\\
\label{eq:Jacobian}
\end{eqnarray}
\renewcommand{\arraystretch}{1}
Now, a small fluctuation $\delta{\bm x}$ follows $\dot{\delta}{\bm x}\simeq{\bm J}\delta{\bm x}$, where $\delta{\bm x}\equiv{\bm x}-{\bm x}_{\rm ss}$ with the steady state values ${\bm x}_{\rm ss}=(x_{\rm ss},y_{\rm ss}, n_{\rm ss}, \theta_{\rm ss})$. We calculate the eigenvalues of ${\bm J}({\bm x})$ at the steady states ${\bm x}={\bm x}_{\rm ss}$ for various input power $P$ and detuning $\delta$. When the pair of the eigenvalues of the Jacobian ${\bm J}({\bm x})$ have positive real values, the steady state ${\bm x}_{\rm ss}$ becomes unstable, which leads to limit cycle oscillation (the Hopf bifurcation) \cite{Strogatz2018,Kuramoto2003}. We show nontrivial regions as functions of input power $P$ and detuning $\delta$ in Fig. \ref{fig:schematic_single}(b), where the bistable and limit cycle (self-pulsing) region are indicated by ``BS" and ``SP", respectively. In the SP+BS region, one steady state is stable, while the other is unstable. In this paper, since we are interested in limit cycle oscillation, we focus solely on the SP region. We also note that the Jacobian matrix Eq. (\ref{eq:Jacobian}) is used again in Section 4. Additionally, in Fig. \ref{fig:schematic_single}(c), we show the input power $P$ and detuning $\delta$ dependence of the limit cycle's frequency $\Omega$, which were obtained from numerical time evolutions. Figure \ref{fig:schematic_single}(c) indicates that the limit cycle's frequency decreases with increasing pump power $P$. 

Now, we directly simulate coupled-mode equations (\ref{eq:coupled_alpha})-(\ref{eq:coupled_theta}). The real-time evolutions of light output $I(t)=|\alpha(t)|^2$ (left), carrier $n(t)$ (right), and $\theta(t)$ (right) are shown in Fig. \ref{fig:schematic_single}(d), where the detuning is $\delta=-2\kappa$, and laser input powers are $P=0.6\kappa^2$ (d) and $1.0\kappa^2$ (e). In Fig. \ref{fig:schematic_single}(d), which is for $P=0.6\kappa^2$, all the variables reach steady states when $t\simeq1000$ ns, and there is no self-pulsing. Meanwhile, for $P=1.0\kappa^2$ [see Fig. \ref{fig:schematic_single}(e)], all the variables clearly exhibit temporal periodic oscillations (limit cycle oscillation) with a frequency of $\Omega/2\pi=11$ MHz. In fact, in Fig. \ref{fig:schematic_single}(b), the values $\delta=-2\kappa$ and $P=1.0\kappa^2$ are represented as a blue filled circle in the SP region. Meanwhile, the values $\delta=-2\kappa$ and $P=0.6\kappa^2$ are outside the SP region . In the rest of this paper, we show only light output $I(t)=|\alpha(t)|^2$ because it is the only measurable valuable in experiments. 

Finally, we comment on the origin of limit cycle oscillation in Si PhC microcavities. In a minimum model that exhibits limit cycle oscillation, we set $\eta=0$ and $f=0$ in Eqs. (\ref{eq:coupled_alpha}) and (\ref{eq:coupled_theta}), which have only quantitative effects. Furthermore, the exponent of the term $\kappa\xi|\alpha|^4$ in Eq. (\ref{eq:coupled_n}) is not essential, because limit cycle oscillation appears even if this term is replaced with $\kappa\xi|\alpha|^2$. In fact, limit cycle oscillation requires only that the signs of the nonlinear energy shifts be opposite for carrier and thermal components; that the carrier lifetime be comparable to or even shorter than the photon lifetime, with  the thermal lifetime much longer than the photon lifetime; and that $\beta$ be  much smaller than $\xi$, approximately $\beta/\xi\simeq\gamma_\theta/\gamma_n$, to make carrier- and thermal-induced energy-shifts comparable. 
Due to the large time-scale difference and the opposite sign of the nonlinear energy shifts, a delayed positive feedback instantaneously occurs when the effective cavity frequency returns to the frequency of the laser input, which leads to self-pulsing.

\section{Coupled limit cycle dynamics}
\begin{figure}
\centering\includegraphics[width=11cm]{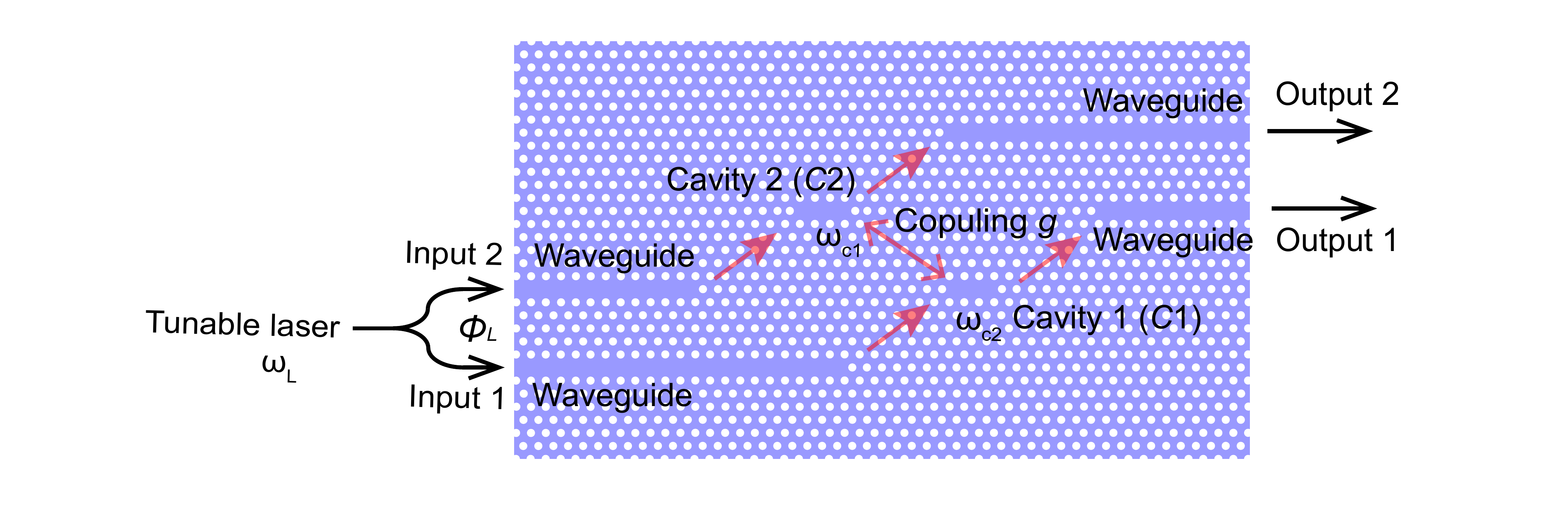}
\caption{Schematic of coupled high-Q Si PhC microcavities. The two cavities are labelled as $C1$ and $C2$. The coherent coupling $g$ is introduced through tunneling of evanescent fields, which is linear coupling. The light outputs from the two cavities are extracted from the two output waveguides, while the two cavities are driven though the two input waveguides. The two laser inputs from a single laser source are separated with on-chip Si wire waveguides and a beam splitter. The phase difference of the laser inputs $\phi_L$ is adjusted by the optical path lengths of the on-chip Si wire waveguides.}
\label{fig:schematic_coupled}
\end{figure}

. The proposed device with two coupled Si PhC cavities is sketched in Fig. \ref{fig:schematic_coupled}, where the two cavities are labelled as $C1$ and $C2$. Since the cavities are evanescently coupled, coupling strength $g$ depends on the distance between the two cavities. To drive the two cavities, we separate a single laser source into two inputs using on-chip Si wire waveguides instead of two laser sources. This process is very important for temporally fixing  the relative phase difference $\phi_L$ between the two laser inputs. Actually, we show that the relative phase difference $\phi_L$ has a crucial impact on synchronization. The design shown in Fig. \ref{fig:schematic_coupled} has two output waveguides, which are used to measure light outputs from $C1$ and $C2$. 

Coupled-mode equations (\ref{eq:coupled_alpha})-(\ref{eq:coupled_theta}) for a single Si PhC cavity are easily extended to the two coupled cavities as
\begin{eqnarray}
\dot{\alpha}_1&=&\kappa_1\lbrace i(-\delta_1/\kappa_1-\theta_1+n_1)-(1+fn_1)\rbrace\alpha_1-ig\alpha_2+\sqrt{P_1}\label{eq:coupled_alpha1}\\
\dot{n}_1&=&-\gamma_1n_1+\kappa_1\xi|\alpha_1|^4\label{eq:coupled_n1}\\
\dot{\theta}_1&=&-\Gamma_1\theta_1+\kappa_1\beta|\alpha_1|^2+\kappa_1\eta|\alpha_1|^2n_1\\\label{eq:coupled_theta1}
&&\nonumber\\
\dot{\alpha}_2&=&\kappa_2\lbrace i(-\delta_2/\kappa_2-\theta_2+n_2)-(1+fn_2)\rbrace\alpha_2-ig\alpha_1+\sqrt{P_2}e^{i\phi_L}\label{eq:coupled_alpha2}\\
\dot{n}_2&=&-\gamma_2n_2+\kappa_2\xi|\alpha_2|^4\label{eq:coupled_n2}\\
\dot{\theta}_2&=&-\Gamma_2\theta_2+\kappa_2\beta|\alpha_2|^2+\kappa_2\eta|\alpha_2|^2n_2,\label{eq:coupled_theta2}
\end{eqnarray}
Equations (\ref{eq:coupled_alpha1})-(\ref{eq:coupled_theta1}) and Eqs. (\ref{eq:coupled_alpha2})-(\ref{eq:coupled_theta2}) represent dynamics for $C1$ and $C2$, respectively. The coherent field coupling between $C1$ and $C2$ is represented by the coupling strength $g$. In Eq. (\ref{eq:coupled_alpha2}), the term $e^{i\phi_L}$ represents a phase factor originating from the phase difference between the two laser inputs. It is worth noting that $\phi_L$ is the phase associated with the field, and thus is not directly related to a limit cycle's phase, which is introduced in Section 4. The values of the nonlinear coefficients $f$, $\xi$, $\beta$, and $\eta$ are the same as those in Fig. \ref{fig:schematic_single}. The cavity detuning is defined as $\delta_{1,2}\equiv\omega_L-\omega_{1,2}$, where $\omega_{1,2}$ is the resonance frequency of the cavity. 

To observe synchronization, there must to be a small frequency difference in two limit cycles. However, in our proposal, the two cavities are designed to be identical because natural disorders or unavoidable fabrication errors will introduce an intrinsic parameter and resonance frequency difference between the two cavities. In this section, for the demonstration of synchronization, we consider a rather ideal device. Namely, only the cavity resonance frequencies are slightly different: $\delta_1=\omega_L-\omega_1=-2\kappa_1$, while $\delta_2=\omega_L-\omega_2=-1.5\kappa_1$. The other parameters are the same for the cavity $C1$ and $C2$: $1/2\kappa_1=1/2\kappa_2=300$ ps, $1/\gamma_1=1/\gamma_2=200$ ps, and $1/\Gamma_1=1/\Gamma_2=100$ ns. Additionally, we derive the two cavities with the same input powers, $P_1=P_2=\kappa_1$. In Section 5, we consider a more realistic device, where the resonance frequencies of the two cavities are moderately different.
\begin{figure}
\centering\includegraphics[width=11cm]{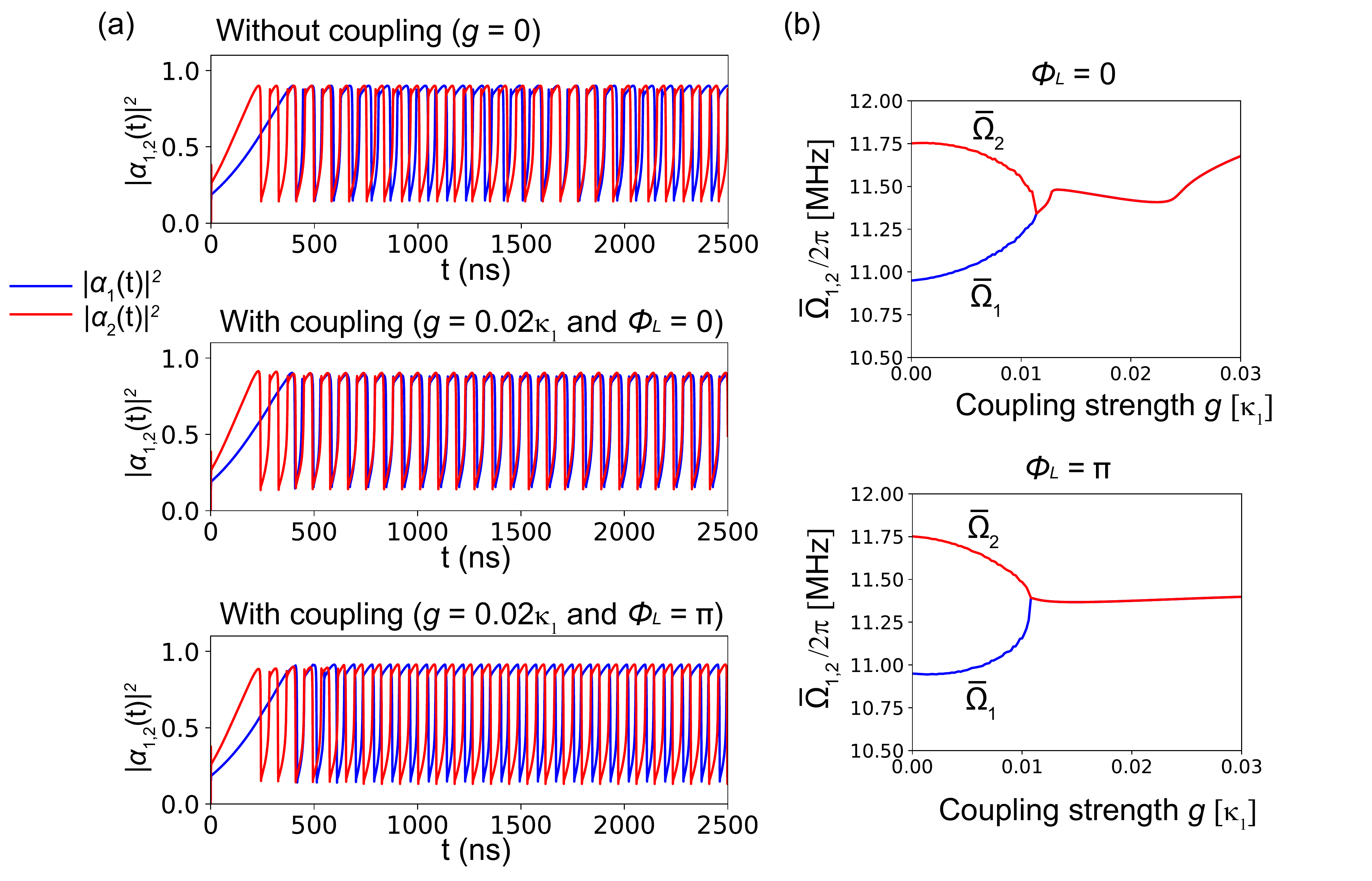}
\caption{(a) Simulated time evolutions of the light output intensity $|\alpha_{1,2}(t)|^2$ without (top) and with coupling $g=0.02\kappa_1$ (middle and bottom). The middle and bottom time evolutions are for $\phi_L=0$ and $\pi$, respectively. (b) The average frequencies $\bar{\Omega}_{1,2}$ of the two limit cycle oscillations $|\alpha_{1,2}(t)|^2$ for $\phi_L=0$ (upper) and $\pi$ (lower) as a function of the coupling strength $g$. The critical coupling strengths of synchronization are $g_c=0.0115$ and $0.011$ for $\phi_L=0$ and $\pi$, respectively. In these simulations, we used $1/2\kappa_1=1/2\kappa_2=300$ ps, $1/\gamma_1=1/\gamma_2=200$ ps, $1/\Gamma_1=1/\Gamma_2=100$ ns, $P_1=P_2=\kappa_1^2$, $\delta_1=\omega_L-\omega_{1}=-2\kappa_1$, and $\delta_2=\omega_L-\omega_{2}=-1.5\kappa_1$.}
\label{fig:sync}
\end{figure}  

Figure \ref{fig:sync}(a) shows time evolutions of the light output $|\alpha_{1,2}(t)|^2$ without $g=0$ (top) and with coupling $g=0.02\kappa_1$ (middle and bottom), which are the central results of this paper. In Fig. \ref{fig:sync}(a), the middle and bottom time evolutions are for the phase difference $\phi_L=0$ and $\phi_L=\pi$, respectively. Note that this coupling strength ($g=0.02\kappa_1$) is much smaller than the cavity decay rates ($g\ll\kappa_{1,2}$), so  the two coupled cavities are in the weak-coupling regime. Without coherent coupling ($g=0$) [Fig. \ref{fig:sync}(a), top], the two limit cycle oscillations are completely decoupled and thus have their own frequencies: $\Omega_1/2\pi=11$ and $\Omega_2/2\pi=11.75$ MHz for $C1$ and $C2$, respectively. On the other hand, when small coupling $g=0.02\kappa_1$ is introduced, the time evolution dramatically changes as shown in the middle and bottom time evolutions in \ref{fig:sync}(a), where the two limit cycle oscillations are perfectly synchronized (entrained) with each other. Furthermore,   we notice that synchronization is in-phase for $\phi_L=0$ (middle), while ``anti-phase" for $\phi_L=\pi$ (bottom). 

Now, we briefly discuss a synchronization time. By turning on coupling for ``steady-state" uncoupled limit cycle oscillations (not shown), we found that the synchronization time for $g=0.02\kappa_1$ is about 500 ns, which corresponds to approximately the five periods of the limit cycle oscillations. In fact, while the oscillation frequency of limit cycles is typically about $\sim$10 MHz in Fig. \ref{fig:sync}(a), the strength of coherent coupling $g=0.02\kappa_1$ ($1/g=1/0.02\kappa_1=30$ ns) corresponds to 5.3 MHz.
We also found that when the coupling strength is increased to $g=0.05\kappa_1$, the synchronization time actually becomes comparable to the period of the limit cycles (not shown). 

In addition to the time evolutions, we show in Fig. \ref{fig:sync}(b) the mean frequencies $\bar{\Omega}_{1,2}$ of the two limit cycles as a function of coupling strength $g$, where the upper and lower graphs are for $\phi_L=0$ and $\pi$, respectively.  We used the mean frequencies because the oscillations are not perfectly periodic when the coupling strength is smaller than a critical value for synchronization. Figure \ref{fig:sync}(b) clearly shows that as the coupling strength $g$ increases, the mean frequencies $\bar{\Omega}_1$ and $\bar{\Omega}_2$ approach each other and merge when $g$ reaches the critical value $g_c$: $g_c=0.0115\kappa_1$ and $0.011\kappa_1$ for $\phi_L=0$ and $\pi$, respectively. Furthermore, for both $\phi_L=0$ and $\pi$, the frequencies of the synchronized limit cycles are the same: $\Omega_1=\Omega_2$, which is called 1:1 synchronization. Finally, we comment on the fact that the value of $g_c$ is not the same for $\phi_L=0$ and $\pi$. We found that when the frequency difference between uncoupled limit cycles becomes smaller, synchronization occurs with the same critical values of $g_c$ for both $\phi_L=0$ and $\pi$, respectively.

In Appendix A, we show a simulation of an intermediate phase difference $\phi_L=0.5\pi$, which does not exhibit synchronization with coupling strength $g=0.02\kappa_1$.
Furthermore, we discuss synchronization in a near-strong coupling region ($g=\kappa_{1,2}$) in Appendix B. Compared to the very small coupling $g=0.02\kappa_1$ considered in this section, the near-strong coupling region may be technically easy to realize. 

\section{Phase description}  
In Section 3, we demonstrated synchronization of limit cycle oscillations in two cavities by directly simulating time evolutions. For a qualitative understanding of the synchronization, phase reduction theory provides a powerful tool called the phase coupling function \cite{Stankovski2017,Kuramoto2003,Nakao2016}. In particular, the phase coupling function can explain why the in- or anti-phase synchronization occurs depending on the phase difference between the two laser inputs. In this section, after a brief review of phase reduction theory, we numerically derive the phase equation of motion and phase coupling function for coupled-mode equations (\ref{eq:coupled_alpha1})-(\ref{eq:coupled_theta2}).

\subsection{General phase description for a single limit cycle}
The key idea in phase reduction theory is to describe limit cycle dynamics solely with a generalized phase degree of freedom. First, we consider the phase description for general single limit cycle dynamics and introduce a scalar ``phase field" $\phi({\bm x})$. Let us consider a general dynamical system that exhibits limit cycle oscillation:
\begin{equation}
\dot{\bm x}={\bm f}({\bm x}),\label{eq:dynamical_general}
\end{equation}
where ${\bm f}({\bm x})$ is a general function. In the phase description, the phase field $\phi({\bm x})$ is defined in such a way that
\begin{equation}
\dot{\phi}({\bm x})=\nabla_{\bm x}\phi({\bm x})\cdot{\bm f}({\bm x})=\Omega,\label{eq:dynamical_perturbed}
\end{equation}
where $\Omega$ is the frequency of the limit cycle oscillation. If there is no perturbation, dynamics converge on the orbit of the limit cycle and follow the very simple equation of motion $\dot{\phi}=\Omega$, where $\phi$ without any argument represents the phase on the limit cycle's orbit. For simplicity, we denote its orbit as $\bm{\chi}(\phi)$. When the dynamical system [Eq. (\ref{eq:dynamical_general})] is perturbed by a force ${\bm p}({\bm x})$ as $\dot{\bm x}={\bm f}({\bm x})+{\bm p}({\bm x},t)$, equation of motion (\ref{eq:dynamical_perturbed}) is modified as 
\begin{equation}
\dot{\phi}({\bm x})=\Omega+\nabla_{\bm x}\phi({\bm x})\cdot{\bm p}({\bm x},t).
\label{eq:dynamical_perturbed2}
\end{equation}
If the perturbation ${\bm p}({\bm x})$ is sufficiently weak, ${\bm x}$ is approximated as a point on the limit cycle's orbit, ${\bm x}\simeq{\bm \chi}(\phi)$. With this approximation, Eq. (\ref{eq:dynamical_perturbed2}) is further simplified as 
\begin{equation}
\dot{\phi}=\Omega+{\bm Z}(\phi)\cdot{\bm P}(\phi,t),\label{eq:phase_eqm}
\end{equation}
where ${\bm Z}(\phi)\equiv\nabla_{{\bm x}=\bm {\bm \chi}(\phi)}\phi({\bm x})$ is called ``sensitivity" \cite{Winfree1967}. Here the capital $P(\phi,t)$ is defined as $P(\phi,t)\equiv{\bm p}({\bm \chi}(\phi),t)$. Equation (\ref{eq:phase_eqm}) is called the phase equation of motion, and it plays a central role in phase reduction theory. Actually, with Eq. (\ref{eq:phase_eqm}), the perturbed limit cycle dynamics are described solely by the phase degree of freedom $\phi$.

Therefore, our next step is to numerically determine the sensitivity ${\bm Z}(\phi)$ for our dynamical system described by coupled-mode equations (\ref{eq:coupled_alpha})-(\ref{eq:coupled_theta}). Fortunately, to numerically obtain ${\bm Z}(\phi)$,  we can use the adjoint method \cite{Ermentrout1996,Nakao2016}, which employs the fact that  ${\bm Z}(\phi)$ satisfies the following equation of motion:
\begin{equation}
\frac{d{\bm Z}(\Omega t)}{dt}=-{\bm J}^\top({\bm \chi}(\Omega t)){\bm Z}(\Omega t),
\label{eq:adjoint}
\end{equation}
where ${\bm J}^\top({\bm \chi}(\Omega t))$ is the transpose of the Jacobian of a dynamical system. In our case, the Jacobian matrix ${\bm J}$ is already given in Eq. (\ref{eq:Jacobian}). Since Eq. (\ref{eq:adjoint}) is unstable for forward time integration due to the minus sign before ${\bm J}^\top$, we need to perform backward time integration as ${d{\bm Z}(-\Omega t')}/{dt'}={\bm J}^\top({\bm \chi}(-\Omega t')){\bm Z}(-\Omega t')$ with $t'=-t$. Additionally, the numerically obtained ${\bm Z}(\phi)$ was normalized as ${\bm Z}(\phi)\cdot {\bm f}({\bm \chi}(\phi))=\Omega$, which is equivalent to Eq. (\ref{eq:dynamical_perturbed}).
Figure \ref{fig:Gamma}(a) shows numerically obtained ${Z}_i(\phi)$, where the index $i$ represents $x$, $y$, $n$, and $\theta$. Importantly, parameters used for calculating ${\bm Z}(\phi)$ are the same as those in Fig. \ref{fig:schematic_single}(e). From Fig. \ref{fig:Gamma}(a), we notice that there is a scale difference between the four components and the shape of ${\bm Z}(\phi)$ is very complicated compared with, for example, the sensitivity of the simple Stuart-Landau model \cite{Nakao2016}.   

\subsection{Phase coupling function}
Here, we extend phase equation of motion (\ref{eq:phase_eqm}) to two coupled  limit cycles. Let us consider two weakly coupled dynamical systems,  both of which exhibit limit cycle oscillations:
\begin{eqnarray}
\dot{{\bm x}_1}&=&{\bm f}({\bm x}_1)+\delta{\bm f}_1({\bm x}_1)+{\bm g}_{12}({\bm x}_1,{\bm x}_2)\label{eq:dynamical_coupled1}\\
\dot{{\bm x}_2}&=&{\bm f}({\bm x}_2)+\delta{\bm f}_2({\bm x}_2)+{\bm g}_{21}({\bm x}_2,{\bm x}_1),\label{eq:dynamical_coupled2}
\end{eqnarray}
where $\delta f_{1,2}({\bm x}_{1,2})$ is a deviation from the ``standard" oscillator ${\bm f}({\bm x})$ [Eq. (\ref{eq:dynamical_general})], while ${\bm g}_{12}({\bm x}_1,{\bm x}_2)$ and ${\bm g}_{21}({\bm x}_2,{\bm x}_1)$ represent coupling between the two systems. Rewriting with the phase coordinate $\phi_{1,2}$ of the standard oscillator and taking the terms $\delta{\bm f}_i({\bm x}_i)$, ${\bm g}_{12}({\bm x}_1,{\bm x}_2)$, and ${\bm g}_{21}({\bm x}_2,{\bm x}_1)$ as perturbations, the phase equations of motion corresponding to Eqs (\ref{eq:dynamical_coupled1}) and (\ref{eq:dynamical_coupled2}) are given by
\begin{eqnarray}
\dot{\phi}_1&=&\Omega+{\bm Z}(\phi_1)\cdot\delta{\bm F}_1(\phi_1)+{\bm Z}(\phi_1)\cdot{\bm G}_{12}(\phi_1,\phi_2)\label{eq:phi1}\\
\dot{\phi}_2&=&\Omega+{\bm Z}(\phi_2)\cdot\delta{\bm F}_2(\phi_2)+{\bm Z}(\phi_2)\cdot{\bm G}_{21}(\phi_2,\phi_1),\label{eq:phi2}
\end{eqnarray}
where the upper-case symbols represent the functions of the standard oscillator's phase $\phi_{1,2}$, which is given by $\dot{\bm x}_{1,2}={\bm f}({\bm x}_{1,2})$. For further simplification of Eqs. (\ref{eq:phi1}) and (\ref{eq:phi2}), we transform $\phi_{1,2}$ into the rotating frame of the standard oscillator as $\psi_{1,2}\equiv\phi_{1,2}-\Omega t$, where $\Omega$ is the standard oscillator' oscillation frequency. Additionally, we perform an approximation for the coupled phase equations of motion by averaging over one period of the standard oscillator. With these procedures, Eqs. (\ref{eq:phi1}) and (\ref{eq:phi2}) become
\begin{eqnarray}
\dot{\psi_1}&=&\delta\Omega_1+\Gamma_{12}(\psi_1-\psi_2)\\
\dot{\psi_2}&=&\delta\Omega_2+\Gamma_{21}(\psi_2-\psi_1).
\end{eqnarray}
Here, the frequency shift $\delta\Omega_{1,2}$ and the phase coupling function $\Gamma_{ij}(\psi)$ are given by
\begin{equation}
\delta\Omega_{1,2}=\frac{1}{2\pi}\int_0^{2\pi}d\theta{\bm Z}(\theta)\cdot\delta{\bm F}_{1,2}(\theta)
\end{equation}
and
\begin{equation}
\Gamma_{ij}(\psi)=\frac{1}{2\pi}\int_0^{2\pi}d\eta{\bm Z}(\eta+\psi)\cdot{\bm G}_{ij}(\eta+\psi,\eta),\label{eq:Gamma}
\end{equation}
respectively. Finally, the phase difference between the two oscillators, $\psi=\psi_2-\psi_1$, follows the following simple equation:
\begin{equation}
\dot{\psi}=\Delta\Omega+\Gamma_a(\psi),\label{eq:phase_eq_final}
\end{equation} 
where $\Delta\Omega\equiv\delta\Omega_2-\delta\Omega_1$ and $\Gamma_a(\psi)\equiv\Gamma_{21}(\psi)-\Gamma_{12}(-\psi)$. In fact, $\Gamma_a(\psi)$ is the anti-symmetric part of the phase coupling function. A synchronization phase $\psi_{\rm sync}$ is required to satisfy $\Gamma_a(\psi_{\rm sync})=0$ and $\Gamma^\prime_a(\psi_{\rm sync})<0$, where the prime represents the derivative. For example, if $\Gamma_a(0)=0$ and $\Gamma^\prime_a(0)<0$, the phase difference $\psi$ is locked to $\psi=0$ by negative feedback, which is in-phase synchronization. Therefore, the shapes of the phase coupling function allow an intuitive interpretation of a synchronization phase. 

\subsection{Phase coupling function for limit cycles in coupled Si PhC cavities}
\begin{figure}
\centering\includegraphics[width=11cm]{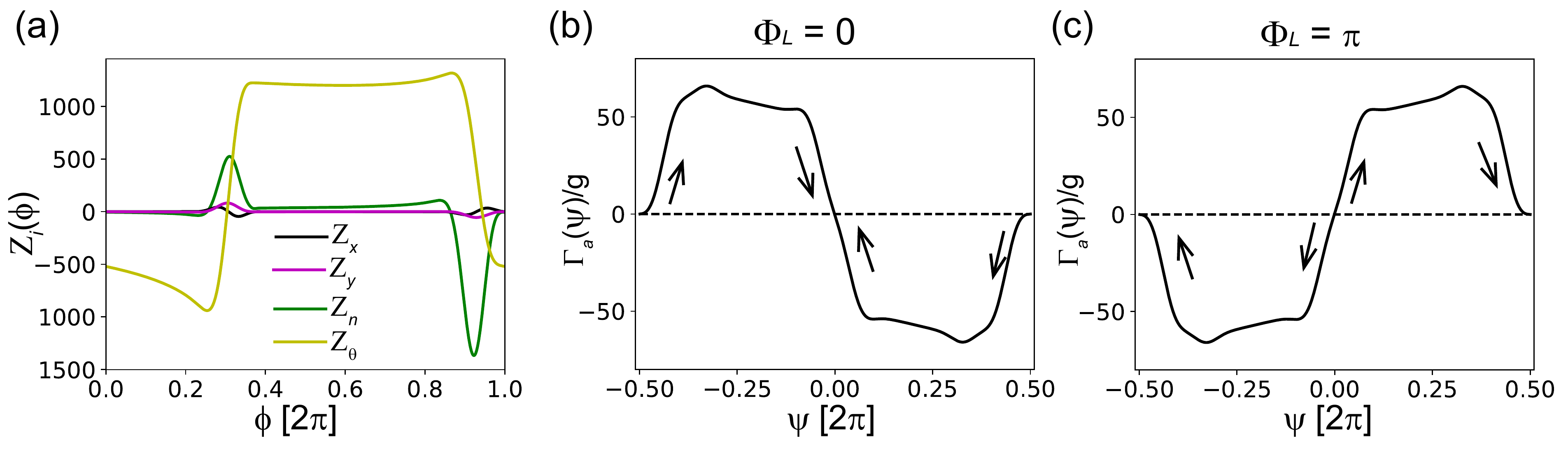}
\caption{(a) Numerically obtained sensitivity ${\bm Z}(\phi)$ for coupled-mode equations (\ref{eq:coupled_alpha})-(\ref{eq:coupled_theta}). (b) Anti-symmetric parts of calculated phase coupling functions $\Gamma_a(\psi)\equiv\Gamma(\psi)-\Gamma(-\psi))$ as a function of the phase $\psi\equiv\psi_2-\psi_1$ for $\phi_L=0$ (left) and $\pi$ (right). The arrows indicates phase locking points. To calculate ${\bm Z}(\phi)$ and $\Gamma_a(\psi)$, we used the same parameter values as those in Fig. \ref{fig:schematic_single} (e).}
\label{fig:Gamma}
\end{figure}

Now, we attempt to numerically calculate the phase coupling function for our dynamical system described by Eqs. (\ref{eq:coupled_alpha1})-(\ref{eq:coupled_theta2}). For this purpose, it is convenient to perform phase rotation for the variable $\alpha_2$ in Eq. (\ref{eq:phi1}) as $\alpha_2e^{-i\phi_L}\rightarrow\alpha_2$. After the phase rotation, Eqs. (\ref{eq:coupled_alpha1})-(\ref{eq:coupled_theta2}) become 
\begin{eqnarray}
\dot{\alpha}_1&=&\kappa_1\lbrace i(-\delta_1/\kappa_1-\theta_1+n_1)-(1+fn_1)\rbrace\alpha_1-ig\alpha_2e^{i\phi_L}+\sqrt{P_1}\label{eq:trans_alpha1}\\
\dot{n}_1&=&-\gamma_1n_1+\kappa_1\xi|\alpha_1|^4\\
\dot{\theta}_1&=&-\Gamma_1\theta_1+\kappa_1\beta|\alpha_1|^2+\kappa_1\eta|\alpha_1|^2n_1\label{eq:trans_theta1}\\
&&\nonumber\\
\dot{\alpha}_2&=&\kappa_2\lbrace i(-\delta_2/\kappa_2-\theta_2+n_2)-(1+fn_2)\rbrace\alpha_2-ig\alpha_1e^{-i\phi_L}+\sqrt{P_2}\label{eq:trans_alpha2}\\
\dot{n}_2&=&-\gamma_2n_2+\kappa_2\xi|\alpha_2|^4\\
\dot{\theta}_2&=&-\Gamma_2\theta_2+\kappa_2\beta|\alpha_2|^2+\kappa_2\eta|\alpha_2|^2n_2,\label{eq:trans_theta2}
\end{eqnarray}
As Eqs. (\ref{eq:trans_alpha1}) and (\ref{eq:trans_alpha2}) indicate, with this transformation, the phase difference between the two lasers $\phi_L$ appears as a ``coupling phase": $-ig\alpha_2\rightarrow -ige^{i\phi_L}\alpha_2$ and $-ig\alpha_1\rightarrow -ige^{-i\phi_L}\alpha_1$. The purpose of this phase rotation is to define a common standard oscillator and its common phase $\phi$ for the two limit cycles.
In fact, except for the coupling terms, Eqs (\ref{eq:trans_alpha1})-(\ref{eq:trans_theta1}) and Eqs (\ref{eq:trans_alpha2})-(\ref{eq:trans_theta2}) are the same equations of motion represented as ${\bm f}({\bm x})$ [Eq. (\ref{eq:stanard_f})].
Meanwhile, the coupling function ${\bm g}_{12}({\bm x}_1,{\bm x}_2)$ and ${\bm g}_{21}({\bm x}_2,{\bm x}_1)$ are given by
\begin{eqnarray}
{\bm g}_{12}({\bm x}_1,{\bm x}_2)=
{\bm g}_{12}({\bm x}_2)=g
\left( \begin{array}{c}
 x_2\sin\phi_L+y_2\cos\phi_L \\
-x_2\cos\phi_L+y_2\sin\phi_L\\
0\\
0
\end{array} \right)\label{eq:g12}
\end{eqnarray} 
and
\begin{eqnarray}
{\bm g}_{21}({\bm x}_2,{\bm x}_1)=
{\bm g}_{21}({\bm x}_1)=g
\left( \begin{array}{c}
x_1\sin(-\phi_L) +y_1\cos(-\phi_L) \\
-x_1\cos(-\phi_L)+y_1\sin(-\phi_L)\\
0\\
0
\end{array} \right),\label{eq:g21}
\end{eqnarray} 
respectively. Additionally, for simplicity, we use the limit cycle in the cavity $C1$ as a standard oscillator, and thus we put $\delta\Omega_1=0$. Since the parameters for the standard oscillator are the same as those in Fig. \ref{fig:schematic_single}(e), we can use the sensitivity ${\bm Z}$ shown in Fig. \ref{fig:Gamma}(a). Representing the coupling function ${\bm g}_{ij}({\bm x}_j)$ with the standard oscillator's phase coordinate as ${\bm G}_{ij}(\phi_j)$, we numerically integrate Eq. (\ref{eq:Gamma}). Figure \ref{fig:Gamma}(b) and (c) show the anti-symmetric parts of the phase coupling function $\Gamma_a(\psi)\equiv\Gamma_{21}(\psi)-\Gamma_{12}(-\psi)$ for $\phi_L=0$ and $\pi$, respectively. Here, the power of the phase description is that the complex limit cycle dynamics represented by coupled-mode equations (\ref{eq:coupled_alpha1})-(\ref{eq:coupled_theta2}) are reduced to a simple phase equation of motion (\ref{eq:phase_eq_final}). In fact, the origin of synchronization is understood only in this phase coordinate. Figure \ref{fig:Gamma}(b) and (c) clearly indicate that when $\phi_L=0$ (b), $\Gamma_a(0)=0$ and $\Gamma^\prime_a(0)<0$ hold, and thus in-phase locking occurs. Meanwhile when $\phi_L=\pi$ (c), $\Gamma_a(\pi)=0$ and $\Gamma^\prime_a(\pi)<0$ hold, and thus anti-phase locking occurs. Here, Fig. \ref{fig:Gamma}(c) is a mirror image of Fig. \ref{fig:Gamma}(b) about the x-axis, which is intuitive because the signs of Eqs (\ref{eq:g12}) and (\ref{eq:g21}) are opposite for $\phi_L=0$ and $\pi$. In our case, since the phase coupling function for $\phi_L=0$ [see Fig. \ref{fig:Gamma}(b)] resembles the sine function, in- and anti-phase synchronizations will occur for $\phi_L=0$ and $\pi$, respectively. The surprise is that although the two cavities are in the weak-coupling regime ($g\ll \kappa_{1,2}$), the phase $\phi_L$ in Eqs (\ref{eq:g12}) and (\ref{eq:g21}) strongly modifies synchronization behavior. In fact, since coherent coupling between fields has a (relative) phase degree of freedom, in a coupled-cavity system, it is always important to take the phase into account. 

The parameters used for calculating phase coupling functions in Fig \ref{fig:Gamma}(b) and (c) are again the same as those in Fig. \ref{fig:schematic_single}(e). Note that the shape of the phase coupling function depends on parameters used for phase reduction. For example, if we perform phase reduction with the parameters for the limit cycle in $C2$ shown in Fig. \ref{fig:sync}(a), although the detailed shape of the phase coupling function changes from that in Fig. \ref{fig:Gamma}(b) (not shown), both have qualitatively the same quasi-sinusoidal shapes.

Thus, anti-phase synchronization for $\phi_L=\pi$ is not a general result, which depends on models and parameters. Meanwhile, we found that in-phase synchronization for $\phi_L=0$ seems to be general.

Finally, we discuss why the linear coherent coupling ${\bm g}_{ij}({\bm x}_j)$ gives rise to the nonlinear phase coupling function $\Gamma_{1,j}(\phi_j)$ shown in Fig. \ref{fig:Gamma}. The mathematical answer is the transformation of the coordinate from the Cartesian coordinates ${\bm x}$ into the phase coordinate of the limit cycle $\phi$. Namely, on the phase coordinate, the linear coupling ${\bm g}_{ij}({\bm x}_j)$ appears as a nonlinear function ${\bm G}_{ij}(\phi_j)$. Since limit cycle oscillation itself originates in a nonlinear dissipative system, the transformation from the Cartesian to the phase coordinate is also nonlinear. We can also interpret our synchronization phenomenon as analogous to injection locking \cite{Siegman1986} or mutual injection locking \cite{Kurtz2005} in laser physics. In injection locking, coupling between slave and master lasers is usually provided by partially transmitting mirrors, which is definitely linear coupling. Therefore, although the coupling itself is linear, synchronization occurs with the modulation of the slave laser's field by the master laser. Similarly to injection locking, in our system, the coherent coupling $g$ allows the oscillating light in the cavity $C1$ to modulate the light in the cavity $C2$. Thus, synchronization is interpreted as a response of the limit cycle in the cavity $C2$ ($C1$) to the modulation from C1 ($C2$). 

\section{Synchronization of two moderately different limit cycles}
Until now, we have considered synchronization in rather ideal systems, where the two limit cycles are almost identical and only their cavity resonance frequencies are slightly different. Thus, it is still questionable whether or not realistic Si PhC cavity devices are able to exhibit synchronization of limit cycle oscillations. Even with state-of-the-art fabrication technology, fabrication errors or natural disorders cause, for example, unavoidable resonance frequency differences in cavities. Therefore, in this section, we consider a more realistic device, where two cavities have a moderate resonance frequency difference.
\begin{figure}
\centering\includegraphics[width=11cm]{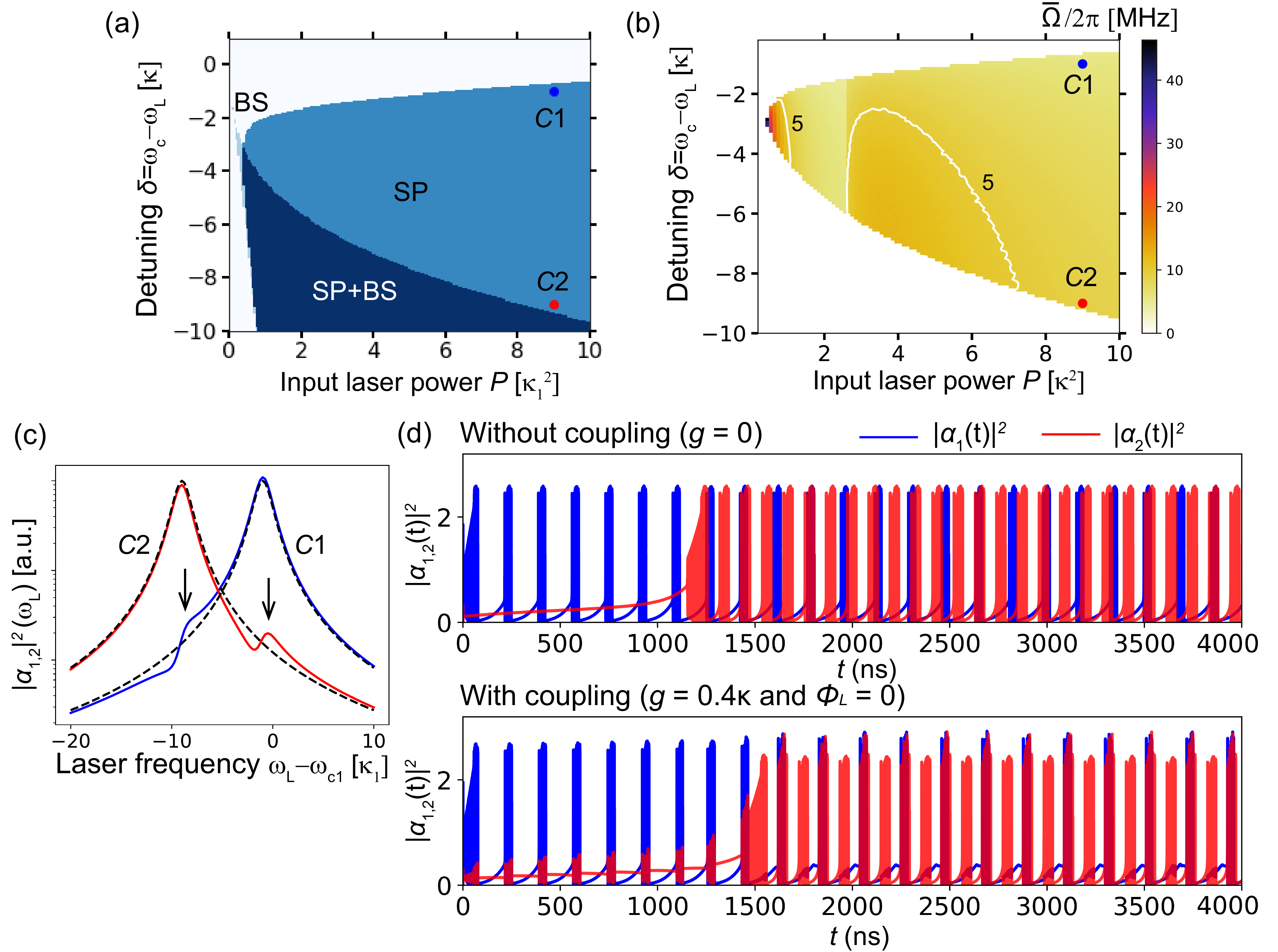}
\caption{Synchronization in realistic cavities with photon lifetime $1/2\kappa_1=1/2\kappa_2=100$ ps and moderate resonance frequency difference $\omega_{2}-\omega_{1}=7\kappa_1$. (a) Self-pulsing (SP) and bistable (BS) regions as functions of laser input power $P$ and detuning $\delta=\omega_L-\omega_c$. (b) Input power $P$ and detuning $\delta$ dependence of the limit cycle's frequency $\Omega$. In (a) and (b), the blue and red filled circles represent parameters used for the cavity $C1$ and $C2$, respectively. (c) Transmission spectra of the coupled cavities obtained as steady state light output intensities $|\alpha_{1,2}(\omega_L)|^2$ as a function of laser input frequency $\omega_L$. The spectra were obtained with the laser input power fixed as $P_1=P_2=0.001\kappa_1^2$. (d) Simulated time evolutions of the light output intensity $|\alpha_{1,2}(t)|^2$ without $g=0$ (upper) and with coupling $g=0.02\kappa_1$ (lower). To simulate the time evolutions, we used $P_1=P_2=9\kappa_1^2$, $\delta_1\equiv\omega_L-\omega_{1}=-1.0\kappa_1$, and $\delta_2\equiv\omega_L-\omega_{2}=-8.0\kappa_1$.}
\label{fig:exp}
\end{figure} 

First, we set photon lifetimes for the two cavities as $1/2\kappa_1=1/2\kappa_2=100$ ps, which correspond to $Q\sim1.0\times10^5$. Importantly, compared with the simulations in previous sections, we slightly decreased the photon lifetime from 300 to 100 ps. This is because it is technically easier to reduce the difference in cavity resonance frequencies for a shorter photon lifetime (a lower $Q$ value). We use the same values as in previous sections for the nonlinear coefficients: $f=0.0244$, $\xi=8.2$, $\beta=0.0296$, and $\eta=0.0036$. For these parameters, the SP and BS regions are represented by the diagram shown in Fig. \ref{fig:exp}(a). Additionally, we show the detuning and input power dependence of the limit cycles' frequency $\Omega$ in Fig. \ref{fig:exp}(b), which is more complicated than Fig. \ref{fig:schematic_single}(c). In fact, the oscillation frequency does not monotonically decrease with increasing input power, because there is an increase in the oscillation frequency at $P\simeq2.6\kappa^2$, and this jump might be related to the onset of fast photon-carrier oscillation \cite{Cazier2013}. Second, we introduce a moderate difference to the cavity resonance frequencies as $\omega_{2}-\omega_{1}=7\kappa_1$. Finally, we also set the value of the coupling strength as $g=0.4\kappa_1$, which is much stronger than in Section 3.

We show the spectra of the two cavities in Fig. \ref{fig:exp}(c), which was obtained by sweeping the laser frequency $\omega_L$ from $\omega_{1}-20\kappa_1$ to $\omega_{1}+20\kappa_1$ and plotting the steady state outputs $|\alpha_1|^2$ and $|\alpha_2|^2$ with very low input power $P_1=P_2=0.001\kappa_1^2$ so as not to induce any nonlinearity. In Fig. \ref{fig:exp}(c), the dashed curves are the spectra without coupling, $g=0$; the solid blue and red curves are the spectra with coupling, $g=0.4\kappa_1$. Comparing the spectra with and without coupling, we notice that the moderately large coupling strength ($g=0.4\kappa$) induces the signatures of coupling as peaks  [see the two arrows in Fig. \ref{fig:exp}(c)] in the spectral tails, but does give rise to normal-mode splitting. Therefore, the system is still in the weak-coupling regime, and we are able to consider coupling as perturbation. 
Here, though the coupling is moderately strong, the system is in weak-coupling because of the large frequency difference between the two cavities $\Delta\omega\equiv\omega_{2}-\omega_{1}=7\kappa_1$. Recall that strong-coupling requires not only $g>\kappa_1$ but also $g>|\Delta\omega|$.
Note also that this value of the resonance frequency difference $\Delta\omega=7\kappa_1$ is experimentally available with state-of-the-art fabrication technology \cite{Notomi2008,Haddadi2014}. Additionally, in Appendix C, we briefly discuss the configuration of two PhC cavities to realize the coupling strength $g=0.4\kappa_1$ with finite-difference time-domain (FDTD) simulations.  

To drive the cavities, we set the detuning values between the cavity resonance and laser frequency as $\delta_1=\omega_{1}-\omega_{L}=-1.0\kappa_1$, which leads to $\delta_2=\omega_{2}-\omega_{L}=-8.0\kappa_1$. Both cavities are driven by inputs with the same power $P_1=P_2=9.0\kappa_1^2$. These parameters are represented by the blue ($C1$) and red ($C2$) filled circles in the diagram in Fig. \ref{fig:exp}(a) and (b), which indicate that both cavities exhibit self-pulsing (limit cycle oscillation). 

Now, in the same way as in Fig. \ref{fig:sync}(a), we show the time evolution of light output intensity $|\alpha_{1,2}(t)|^2$ in Fig. \ref{fig:exp}(d) with (upper) and without coupling (lower). First, we discuss time evolution without coupling, $g=0$. Without coupling, both cavities exhibit limit cycle oscillations with their own frequencies: $\Omega_1/2\pi=5.8$ and $\Omega_2/2\pi=9.2$ MHz for the cavity $C1$ and $C2$, respectively. On the other hand, with coupling, $g=0.4\kappa_1$,  the time evolution clearly shows synchronization of the two oscillations. However, the profile of synchronized oscillations is very different from that in Fig. \ref{fig:sync}(a). For instance, the profile of $|\alpha_{2}(t)|^2$ is strongly modified by introducing the large coupling.

In the synchronized state [see the lower time evolutions in Fig. \ref{fig:exp}(d)], the oscillation periods of the limit cycle oscillations for $|\alpha_{1}(t)|^2$ and $|\alpha_{2}(t)|^2$ are identified as $T_1=T_2=210$ ns, which corresponds to $\Omega_1/2\pi=\Omega_2/2\pi=4.77$ MHz. 
However, from $|\alpha_{2}(t)|^2$ on the lower panel in Fig. \ref{fig:exp}(d), we notice that the limit cycle orbit for $C2$ is strongly modified by coherent coupling compared with the uncoupled orbit. In fact, in terms of the Poincar\'e section, the period of the limit cycle orbit $|\alpha_{2}(t)|^2$ with coupling will be $T_2\simeq105$ ns and the corresponding frequency is $\Omega_2/2\pi\simeq9.54$ MHz, which is the double of 4.77 MHz. This co-existence of the two frequencies and temporal profile for $|\alpha_{2}(t)|^2$  [see Fig. \ref{fig:exp}(d)] may be signatures of period doubling bifurcation \cite{Strogatz2018}, which is an interesting theme for future investigation both from the theoretical and experimental standpoints.

Finally, we comment on the phase difference of the two laser inputs, $\phi_L$. In this section, we have shown the simulation only for $\phi_L=0$ because we found that synchronization occurs only for near-zero phase $\phi_L\simeq0$. Actually, when $\phi_L$ is not close to zero, synchronization does not occur even with $g>\kappa$. This result is related to the large frequency difference between the two uncoupled limit cycles ($\Omega_1=5.8$ and $\Omega_2/2\pi=9.2$ MHz). In fact, if the parameters for the cavity $C1$ and $C2$ are similar , the frequency difference between two uncoupled limit cycles is smaller, and synchronization occurs both for $\phi_L=0$ and $\pi$. Therefore, to realize synchronization in realistic coupled cavities with a moderate frequency difference, it is important to adjust the phase difference of laser inputs to near-zero ($\phi_L\neq0$), which will be achieved by adjusting optical path lengths with, for example, on-chip Si wire waveguides. 

\section{Discussion and future perspective}
First, we argue that the proposed scheme of synchronization is not limited to Si PhC cavities, but applicable to a wide range of limit cycle oscillations in nanophotonic systems, such as nanolasers \cite{Yacomotti2013,Marconi2020}, semiconductor microcavities \cite{Yacomotti2006,Brunstein2012}, and microring resonators \cite{Priem2005,Johnson2006,Pernice2010,VanVaerenbergh2012,Zhang2013}. Actually, the coherent field coupling is easily implemented in these nanophotonic devices, which will lead to synchronization of optical limit cycles. In particular, since coupled-mode equations (\ref{eq:coupled_alpha})-(\ref{eq:coupled_theta}) were originally proposed for modelling optical limit cycles in Si microring resonators \cite{VanVaerenbergh2012,Zhang2013}, our synchronization scheme is easily applicable to them. In terms of the tunability of various physical parameters such as resonance frequencies, Si microring resonators may be advantageous over PhC structures. In particular, a different type synchronization dynamics was investigated with coupled microring resonators in Ref \cite{Xu2019}.
\begin{figure}[h]
\centering\includegraphics[width=11cm]{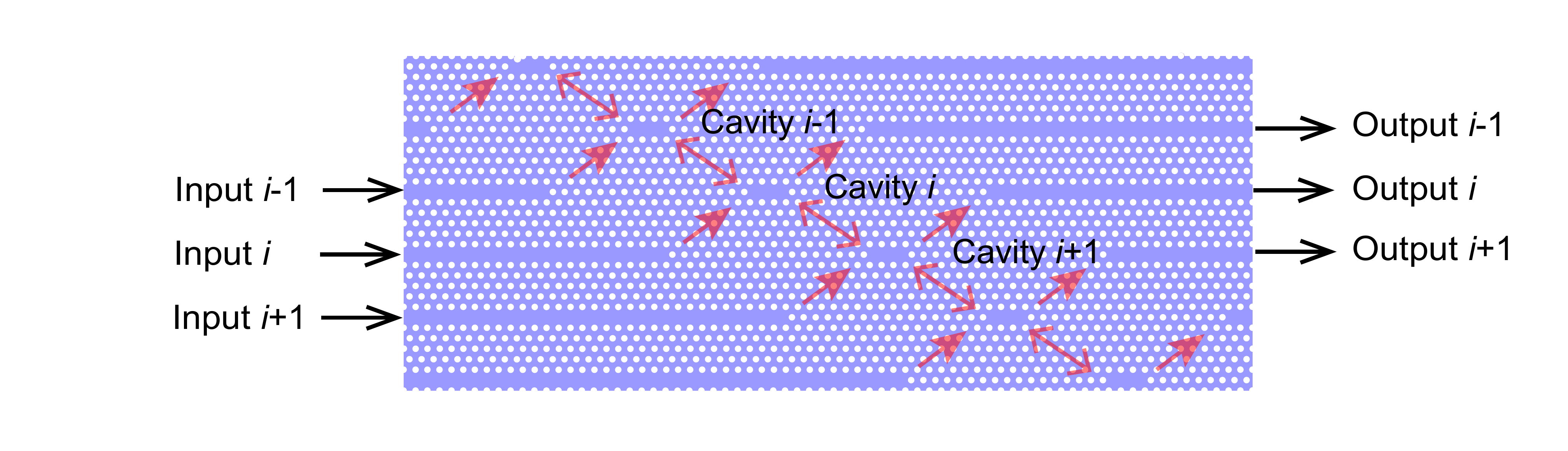}
\caption{Illustration  of an array of coupled Si PhC cavities, which will work as a one-dimensional nearest-neighbor coupling (local) Kuramoto oscillator.}
\label{fig:future}
\end{figure}

Second, we discuss a future perspective of limit cycle synchronization in Si PhC cavities. One can naturally imagine the extension of the two coupled Si PhC cavities to an array of coupled cavities as illustrated in Fig. \ref{fig:future}. In principle, the coupled PhC cavity array illustrated in Fig. \ref{fig:future} could behave as a one-dimensional (1D) nearest-neighbor coupling (local) Kuramoto oscillator. The 1D local Kuramoto model has been theoretically investigated by numerical simulation \cite{Zheng1998} and renormalization group analysis \cite{Daido1988}, which have predicted various nontrivial collective phenomena, including a synchronization state, a phase slip at the onset of de-synchronization, and coupling-induced chaos. From the standpoint of device application, the predicted chaotic state in the 1D local Kuramoto model could be used for photonic reservoir computing \cite{Duport2012}.

\section{Conclusion}
In conclusion, we have theoretically demonstrated synchronization of optical limit cycles with driven coupled silicon (Si) photonic crystal (PhC) cavities, where limit cycle oscillation emerges from carrier- and thermal-induced nonlinearities. Introducing coherent field coupling between two cavities synchronizes (entrains) two limit cycle oscillations. First, we quantitatively demonstrated synchronization by directly simulating the time evolutions of coupled-mode equations. We found that synchronization phase depends on the phase difference of two laser inputs. Second, the numerically simulated synchronization was qualitatively interpreted in the framework of phase description. In particular, we calculated phase coupling functions, which intuitively explain why the synchronization phase depends on the phase difference between the two laser inputs. Finally, we discussed synchronization in a realistic coupled cavity device, where the resonance frequencies of the two cavities are moderately different. Since our proposed design is perfectly compatible with conventional Si fabrication processes, synchronization of optical limit cycles will be easy to implement in future silicon photonic devices and can be extended to an array of coupled cavities. 

\section*{Acknowledgements}
We thank S. Kita, K. Nozaki, and K. Takata for helpful discussions.

\section*{Disclosures}
The authors declare no conflicts of interest.

\section*{Appendix A: Simulations for $\phi_L=0.5\pi$}
\begin{figure}
\centering\includegraphics[width=11cm]{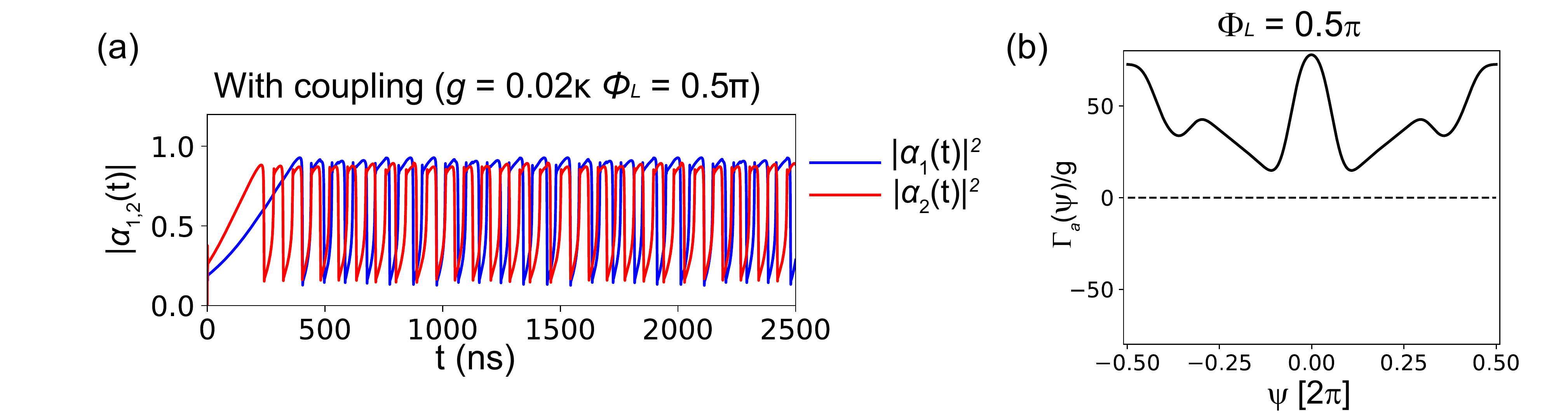}
\caption{(a) Simulated time evolution of the light output intensity $|\alpha_{1,2}(t)|^2$ for $\phi_L=0.5\pi$ with coupling $g=0.02\kappa_1$. (b) Anti-symmetric part of phase coupling function $\Gamma_a(\psi)\equiv\Gamma(\psi)-\Gamma(-\psi)$ as a function of phase $\psi\equiv\psi_2-\psi_1$ for $\phi_L=0.5\pi$. In this figure, all the parameters except for $\phi_L$ are the same as in Fig. \ref{fig:sync}(a).}
\label{fig:half_pi}
\end{figure}

In this appendix, we show simulations when the phase difference in laser inputs is $\phi_L=0.5\pi$ in Fig. \ref{fig:sync}. The simulation described in Section 3 in the main text was performed only for $\phi_L=0$ and $\pi$, which exhibited in- and anti-phase synchronization, respectively. Thus, it is of natural interest to discuss the intermediate case $\phi_L=0.5\pi$. Figure \ref{fig:half_pi}(a) shows the time evolution of light output $|\alpha_{1,2}(t)|^2$ for $\phi_L=0.5\pi$ with coherent coupling $g=0.02\kappa_1$. In fact, in Fig. \ref{fig:half_pi}(a), all the parameters except $\phi_L=$ are the same as in Fig. \ref{fig:sync}(a). Surprisingly, even though the coupling strength is the same as in Fig. \ref{fig:sync}(a), no synchronization is observed in \ref{fig:half_pi}(a). We found that this result can be explained in terms of a phase coupling function. Similarly to \ref{fig:Gamma}(b) and (c) , we show the anti-symmetric part of the phase coupling function in Fig. \ref{fig:half_pi}(b). Interestingly, $\Gamma_a(\psi)$ for $\phi_L=0.5\pi$ never crosses the zero axis, and thus there is no phase locking point. This explains why phase synchronization does not occur for $\phi_L=0.5\pi$ with the small coupling strength ($g=0.02\kappa_1$).

Even for $\phi_L=0.5\pi$, if the coupling strength is further increased, for example, to $g\simeq0.2\kappa_1$, synchronization occurs (not shown). However, this synchronization with a large coupling strength may not be interpreted as 1:1 synchronization, because, there is no smooth transition of the limit cycles' average frequencies from the independent to synchronized state for $\phi_L=0.5\pi$. In summary, for $\phi_L=0.5\pi$, 1:1 synchronization does not occur, but m:n synchronization can occur with a large value of coupling.

\section*{Appendix B: Synchronization in strong coupling regions}
We demonstrate synchronization in the strong coupling region of two coupled cavities. For this purpose, we investigate a near-strong coupling region where $g=\kappa_1=\kappa_2$ and $g>|\Delta\omega|$ with $\Delta\omega\equiv\omega_2-\omega_1=0.5\kappa_1$. 
Although synchronization with weak coherent coupling is theoretically interesting, the very weak coupling strength $g=0.02\kappa_1$ for the high-Q cavities $1/2\kappa_1=1/2\kappa_2=300$ ps used in Section 3 may not be easy in real PhC devices, it is important to discuss coupled limit cycles in the strong coupling region.

\begin{figure}
\centering\includegraphics[width=11cm]{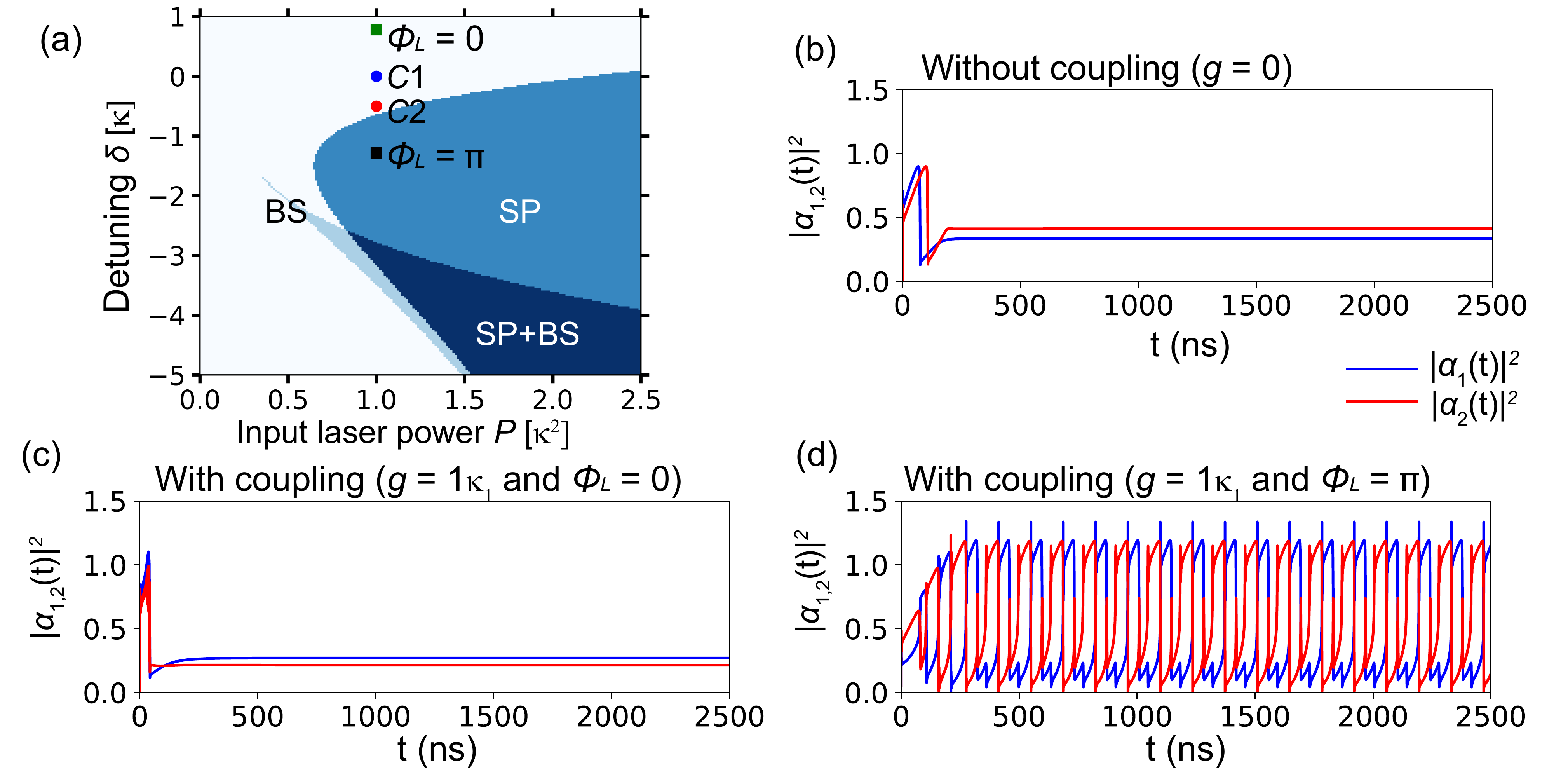}
\caption{(a) Simulations for a near strong-coupling region ($g=\kappa_1=\kappa_2$). For the cavity $C1$ and $C2$, we use $\delta_1=\omega_L-\omega_1=0$ and $\delta_2=\omega_L-\omega_2=-0.5\kappa_1$, respectively. The other parameters are the same as those in Fig. \ref{fig:sync}. (a) Self-pulsing (SP) and bistable (BS) regions as functions of laser input power $P$ and detuning $\delta(=\omega_L-\omega_c)$. The blue and red filled circles represent parameters used for the cavity $C1$ and $C2$, respectively. Meanwhile the green and black filled squares represent the expected parameters for ``bonding" (for $\phi_L=0$) and ``anti-bonding" normal-modes  (for $\phi_L=\pi$) formed by the cavity fields, respectively. (b,c,d) Simulated time evolutions of the light output intensity $|\alpha_{1,2}(t)|^2$ without (b) and with coupling $g=\kappa_1=\kappa_2$ (c,d). The phases are $\phi_L=0$ and $\pi$ for (c) and (d), respectively.}
\label{fig:strong}
\end{figure}

In this Appendix, for the detuning, we set $\delta_1=\omega_L-\omega_1=0$ and $\delta_2=\omega_L-\omega_2=-0.5\kappa_1$ for the cavity $C1$ and $C2$, respectively. The other parameters except for the coupling strength and detuning are the same as those in Fig. \ref{fig:sync}. Thus, for laser input powers, we used $P_1=P_2=\kappa_1^2$. 
Importantly, when there is no coupling ($g=0$), no limit cycle oscillations appear with these parameters as indicated by the time evolutions shown in Fig. \ref{fig:strong}(b). In fact, the parameters for $C1$ and $C2$ are, respectively, indicated by the blue and red filled circles on the trivial region in Fig. \ref{fig:strong}(a).

On the other hand, when a near-strong coupling ($g=\kappa_1$) is introduced, there are still no limit cycle oscillations for $\phi_L=0$, while synchronized limit cycle oscillations, surprisingly, appear for $\phi_L=\pi$. These phase $\phi_L$ dependent results can be understood by considering the normal-modes of the fields $\alpha_1$ and $\alpha_2$. 
When $\phi_L=0$, the electric fields $\alpha_1$ and $\alpha_2$ form a ``bonding" normal-mode. Since the frequency (energy) of the bonding normal-mode $\omega_b$ is lower than the original cavity frequencies $\omega_1$ and $\omega_2$ ($\omega_b<\omega_1,\omega_2$), the corresponding detuning $\delta_a(\equiv\omega_L-\omega_a)$ is still outside the SP region and no limit cycle oscillations appear [see the green filled square in Fig. \ref{fig:strong}(a)].
Meanwhile, when $\phi_L=\pi$, the electric fields $\alpha_1$ and $\alpha_2$ form an ``anti-bonding" normal-mode whose frequency (energy) $\omega_a$ is higher than the original cavity frequencies: $\omega_a>\omega_1,\omega_2$. Therefore, the corresponding detuning $\delta_a(\equiv\omega_L-\omega_a)$ is lower than $\delta_{1,2}$ and enters the SP (self-pulsing) region [see the black filled square in Fig. \ref{fig:strong}(a)].
Note that to plot the green and black filled square in Fig. \ref{fig:strong}(a), we used the frequencies of the bonding $\omega_b$ and anti-bonding normal-modes $\omega_a$ given by
\begin{equation}
\omega_b=\frac{1}{2}\left[\omega_1+\omega_2-\sqrt{(\omega_1-\omega_2)^2+(2g)^2}\right]
\end{equation}
and
\begin{equation}
\omega_a=\frac{1}{2}\left[\omega_1+\omega_2+\sqrt{(\omega_1-\omega_2)^2+(2g)^2}\right],
\end{equation}
respectively.
We also note that, the temporal profile of the synchronized oscillations in Fig. \ref{fig:strong}(d) exhibits anti-phase synchronization, but is quantitatively different from those in Fig. \ref{fig:sync}(a) for $\phi_L=\pi$ due to the large coupling strength $g=\kappa_1$.  

In conclusion, even in the strong-coupling region, synchronization can be realized, but we have to take the effect of normal-mode splitting into account. 
However, note that, the weak-coupling region is more interesting than the strong-coupling region from the standpoint of synchronization physics. This is because synchronization in the strong-coupling region can be interpreted simply as a limit cycle oscillation of a normal-mode that appears in both cavities.

\section*{Appendix C: Estimating coherent coupling strength using FDTD simulations}
Here, we briefly discuss the estimation of a coherent coupling strength $g$ in a realistic coupled PhC cavity structure with three dimensional (3D) finite-difference time-domain (FDTD) simulations. The general idea of the FDTD simulation is as follows. We excite the cavity $C1$ at $t=0$ and probe the time evolution of the energy of the electromagnetic field in the cavity $C2$.
The time evolution of the electromagnetic field energy in the cavity $C2$ exhibits damped oscillation, where the oscillation originates from the coherent coupling (tunneling), while the damping is associated with the finite lifetimes of the cavities.
Therefore, the oscillation frequency of the probed damped oscillation of the electromagnetic field energy in $C2$ corresponds to the coupling strength $g$. Note that as the distance between two cavities increases, the oscillation period becomes longer, which results in a very long computation time. In the 3D FDTD simulations in this Appendix, the maximum time range of time evolution was 1 ns, which already took a few days for computation.   

\begin{figure}
\centering\includegraphics[width=11cm]{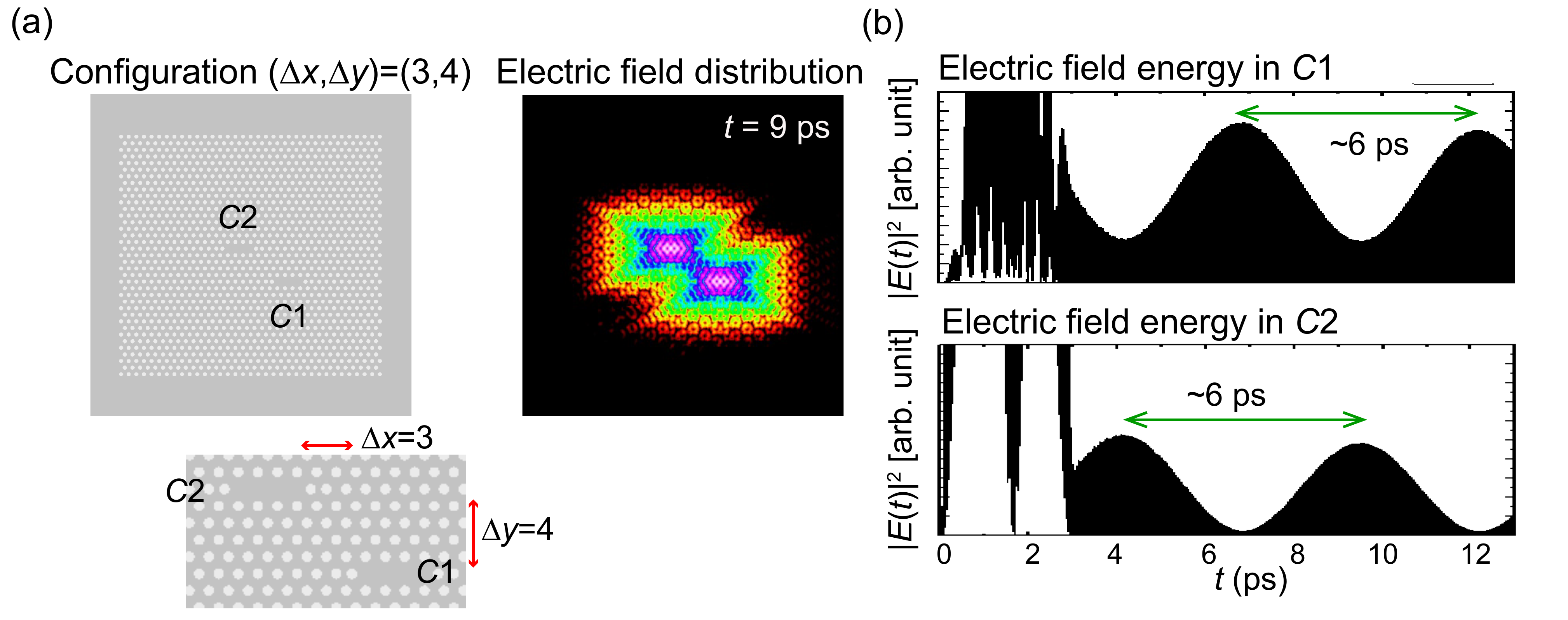}
\caption{(a) Example of a coupled PhC cavity structure and electric field distribution at $t=9$ ps after pulse excitation calculated with 3D FDTD. On the zoomed coupled PhC cavities in (a), the bidirectional arrows $\Delta x=3$ and $\Delta y=4$ represent the $x$- and $y$-direction spacing between $C1$ and $C2$ measured by the number of air-holes. (b) Time evolutions of electric field energies probed in $C1$ and $C2$ after pulse excitation to $C1$.}
\label{fig:FDTD}
\end{figure}

Figure \ref{fig:FDTD}(a) illustrates a coupled PhC cavity structure and simulated electric field distribution at $t=9$ ps after pulse excitation to $C1$, while (b) represents the time evolutions of electric field energy probed in $C1$ and $C2$.
Figure \ref{fig:FDTD}(b) indicates that the oscillation phases of the electric field energies in $C1$ and $C2$ are opposite, which is the evidence of the coherent energy transfer between the two cavities. Furthermore, from the oscillation period of the time evolution in Fig. \ref{fig:FDTD}(b), we can extract the coupling strength as $1/g=6/2\pi$ [ps].

We investigated how the coupling strength changes depending on the configuration of two cavities denoted by $\Delta x$ and $\Delta y$, which represent the $x$- and $y$-direction spacing between the two cavities measured by the number of air-holes, respectively [see the the bidirectional arrows in Fig. \ref{fig:FDTD}(a)].
For example, the configuration of the coupled PhC cavities shown in Fig. \ref{fig:FDTD}(a) is denoted as $(\Delta x,\Delta y)=(3,4)$.
In Table \ref{table1}, we summarized the coupling strength for the five different configurations of two coupled cavities. 
For all 3D FDTD simulations, the total calculation area was 18.4 $\mu$m$\times$18.4 $\mu$m$\times$4 $\mu$m, while the lattice constant was 435 nm. The cavity C2 was on-resonantly excited with a pulse electric field whose pulse width is 3 ps. Depending on the configuration of the two cavities, their cavity lifetime vary from 100 to 200 ps, which must be at least longer than the period of the coherent oscillation ($2\pi/g$).
\def\arraystretch{1.5}%
\begin{table}[htbp]
 \centering
 \begin{tabular}{|c|c|c|c|c|c|}
 \hline 
           $\Delta x=$ & 2 & 3 & 4  & 4 & 7\\ 
           \hline 
           $\Delta y=$ & 2 & 4 & 5  & 6 & 7\\
  \hline
 $\frac{2\pi}{g}\simeq$ & 2 ps& 6 ps& 16 ps& 80 ps& 380 ps\\
 \hline
 \end{tabular}
 \vspace*{3mm}
 \caption{Coupling strength $2\pi/g$ obtained from the FDTD simulation for various configurations denoted by spacing between the two cavities $\Delta x$ and $\Delta y$ [see the the bidirectional arrows in Fig. \ref{fig:FDTD}(a)]. Actually, $2\pi/g$ is the period of coherent oscillation of electromagnetic field energies.}
 \label{table1}
\end{table}
Due to the limited computation time, the weakest coupling strength in Table \ref{table1} was $1/g=380/2\pi\simeq60$ [ps] for the configuration $(\Delta x,\Delta y)=(7,7)$. Compared with the field decay rate $1/\kappa_1=200$ ps assumed in Section 5, this coupling strength $1/g\simeq60$ is still stronger than the field decay rate: $g>\kappa_1$.

Although our FDTD simulations failed to calculate the weak-coupling region for $1/\kappa_1=200$, Table \ref{table1} provides a hint of cavity configuration to realize the weak-coupling region.
In fact, Table \ref{table1} indicates that the coupling strength seems to exponentially decreases with an increase in the distance between the cavities.
Therefore, it is natural to expect that the coupling strength $g=0.4\kappa_1$ with $1/\kappa_1=200$, which is assumed in Section 5, will be soon achieved by slightly increasing the distance between cavities, for example, as $(\Delta x,\Delta y)=(8,8)$.   
Of course, weak coupling is also realized by decreasing the cavity photon lifetime. However, we found that, when the photon lifetime is further decreased, the numerical integration of coupled-mode equations (\ref{eq:coupled_alpha})-(\ref{eq:coupled_theta}) become unstable due to the too large time scale difference between the field, carrier, and thermal components
\cite{Takemura2020}.
Finally, we comment on an alternative strategy proposed in \cite{Haddadi2014}, which is worth considering for the design of coupled PhC cavities. In fact, Ref. \cite{Haddadi2014} demonstrated a barrier engineering technique for robustly tailoring a coupling strength between cavities, which employs the modulation of the radius of air-holes between two cavities.


\begin{thebibliography}{58}%
\makeatletter
\providecommand \@ifxundefined [1]{%
 \@ifx{#1\undefined}
}%
\providecommand \@ifnum [1]{%
 \ifnum #1\expandafter \@firstoftwo
 \else \expandafter \@secondoftwo
 \fi
}%
\providecommand \@ifx [1]{%
 \ifx #1\expandafter \@firstoftwo
 \else \expandafter \@secondoftwo
 \fi
}%
\providecommand \natexlab [1]{#1}%
\providecommand \enquote  [1]{``#1''}%
\providecommand \bibnamefont  [1]{#1}%
\providecommand \bibfnamefont [1]{#1}%
\providecommand \citenamefont [1]{#1}%
\providecommand \href@noop [0]{\@secondoftwo}%
\providecommand \href [0]{\begingroup \@sanitize@url \@href}%
\providecommand \@href[1]{\@@startlink{#1}\@@href}%
\providecommand \@@href[1]{\endgroup#1\@@endlink}%
\providecommand \@sanitize@url [0]{\catcode `\\12\catcode `\$12\catcode
  `\&12\catcode `\#12\catcode `\^12\catcode `\_12\catcode `\%12\relax}%
\providecommand \@@startlink[1]{}%
\providecommand \@@endlink[0]{}%
\providecommand \url  [0]{\begingroup\@sanitize@url \@url }%
\providecommand \@url [1]{\endgroup\@href {#1}{\urlprefix }}%
\providecommand \urlprefix  [0]{URL }%
\providecommand \Eprint [0]{\href }%
\providecommand \doibase [0]{http://dx.doi.org/}%
\providecommand \selectlanguage [0]{\@gobble}%
\providecommand \bibinfo  [0]{\@secondoftwo}%
\providecommand \bibfield  [0]{\@secondoftwo}%
\providecommand \translation [1]{[#1]}%
\providecommand \BibitemOpen [0]{}%
\providecommand \bibitemStop [0]{}%
\providecommand \bibitemNoStop [0]{.\EOS\space}%
\providecommand \EOS [0]{\spacefactor3000\relax}%
\providecommand \BibitemShut  [1]{\csname bibitem#1\endcsname}%
\let\auto@bib@innerbib\@empty
\bibitem [{\citenamefont {Pikovsky}\ \emph {et~al.}(2003)\citenamefont
  {Pikovsky}, \citenamefont {Kurths}, \citenamefont {Rosenblum},\ and\
  \citenamefont {Kurths}}]{Pikovsky2003}%
  \BibitemOpen
  \bibfield  {author} {\bibinfo {author} {\bibfnamefont {A.}~\bibnamefont
  {Pikovsky}}, \bibinfo {author} {\bibfnamefont {J.}~\bibnamefont {Kurths}},
  \bibinfo {author} {\bibfnamefont {M.}~\bibnamefont {Rosenblum}}, \ and\
  \bibinfo {author} {\bibfnamefont {J.}~\bibnamefont {Kurths}},\ }\href@noop {}
  {\emph {\bibinfo {title} {Synchronization: a universal concept in nonlinear
  sciences}}},\ Vol.~\bibinfo {volume} {12}\ (\bibinfo  {publisher} {Cambridge
  university press},\ \bibinfo {year} {2003})\BibitemShut {NoStop}%
\bibitem [{\citenamefont {Appleton}(1922)}]{Appleton1922}%
  \BibitemOpen
  \bibfield  {author} {\bibinfo {author} {\bibfnamefont {E.~V.}\ \bibnamefont
  {Appleton}},\ }\href@noop {} {\enquote {\bibinfo {title} {Automatic
  synchronization of triode oscillators},}\ } (\bibinfo {year}
  {1922})\BibitemShut {NoStop}%
\bibitem [{\citenamefont {Van Der~Pol}(1927)}]{VanDerPol1927}%
  \BibitemOpen
  \bibfield  {author} {\bibinfo {author} {\bibfnamefont {B.}~\bibnamefont {Van
  Der~Pol}},\ }\href@noop {} {\bibfield  {journal} {\bibinfo  {journal} {The
  London, Edinburgh, and Dublin Philosophical Magazine and Journal of Science}\
  }\textbf {\bibinfo {volume} {3}},\ \bibinfo {pages} {65} (\bibinfo {year}
  {1927})}\BibitemShut {NoStop}%
\bibitem [{\citenamefont {Winfree}(1967)}]{Winfree1967}%
  \BibitemOpen
  \bibfield  {author} {\bibinfo {author} {\bibfnamefont {A.~T.}\ \bibnamefont
  {Winfree}},\ }\href {\doibase https://doi.org/10.1016/0022-5193(67)90051-3}
  {\bibfield  {journal} {\bibinfo  {journal} {Journal of Theoretical Biology}\
  }\textbf {\bibinfo {volume} {16}},\ \bibinfo {pages} {15 } (\bibinfo {year}
  {1967})}\BibitemShut {NoStop}%
\bibitem [{\citenamefont {Kuramoto}(2003)}]{Kuramoto2003}%
  \BibitemOpen
  \bibfield  {author} {\bibinfo {author} {\bibfnamefont {Y.}~\bibnamefont
  {Kuramoto}},\ }\href@noop {} {\emph {\bibinfo {title} {Chemical oscillations,
  waves, and turbulence}}}\ (\bibinfo  {publisher} {Courier Corporation},\
  \bibinfo {year} {2003})\BibitemShut {NoStop}%
\bibitem [{\citenamefont {Tsang}\ \emph {et~al.}(1991)\citenamefont {Tsang},
  \citenamefont {Mirollo}, \citenamefont {Strogatz},\ and\ \citenamefont
  {Wiesenfeld}}]{Tsang1991}%
  \BibitemOpen
  \bibfield  {author} {\bibinfo {author} {\bibfnamefont {K.~Y.}\ \bibnamefont
  {Tsang}}, \bibinfo {author} {\bibfnamefont {R.~E.}\ \bibnamefont {Mirollo}},
  \bibinfo {author} {\bibfnamefont {S.~H.}\ \bibnamefont {Strogatz}}, \ and\
  \bibinfo {author} {\bibfnamefont {K.}~\bibnamefont {Wiesenfeld}},\ }\href
  {\doibase https://doi.org/10.1016/0167-2789(91)90054-D} {\bibfield  {journal}
  {\bibinfo  {journal} {Physica D: Nonlinear Phenomena}\ }\textbf {\bibinfo
  {volume} {48}},\ \bibinfo {pages} {102 } (\bibinfo {year}
  {1991})}\BibitemShut {NoStop}%
\bibitem [{\citenamefont {Wiesenfeld}\ \emph {et~al.}(1996)\citenamefont
  {Wiesenfeld}, \citenamefont {Colet},\ and\ \citenamefont
  {Strogatz}}]{Wiesenfeld1996}%
  \BibitemOpen
  \bibfield  {author} {\bibinfo {author} {\bibfnamefont {K.}~\bibnamefont
  {Wiesenfeld}}, \bibinfo {author} {\bibfnamefont {P.}~\bibnamefont {Colet}}, \
  and\ \bibinfo {author} {\bibfnamefont {S.~H.}\ \bibnamefont {Strogatz}},\
  }\href {\doibase 10.1103/PhysRevLett.76.404} {\bibfield  {journal} {\bibinfo
  {journal} {Phys. Rev. Lett.}\ }\textbf {\bibinfo {volume} {76}},\ \bibinfo
  {pages} {404} (\bibinfo {year} {1996})}\BibitemShut {NoStop}%
\bibitem [{\citenamefont {Barbara}\ \emph {et~al.}(1999)\citenamefont
  {Barbara}, \citenamefont {Cawthorne}, \citenamefont {Shitov},\ and\
  \citenamefont {Lobb}}]{Barbara1999}%
  \BibitemOpen
  \bibfield  {author} {\bibinfo {author} {\bibfnamefont {P.}~\bibnamefont
  {Barbara}}, \bibinfo {author} {\bibfnamefont {A.~B.}\ \bibnamefont
  {Cawthorne}}, \bibinfo {author} {\bibfnamefont {S.~V.}\ \bibnamefont
  {Shitov}}, \ and\ \bibinfo {author} {\bibfnamefont {C.~J.}\ \bibnamefont
  {Lobb}},\ }\href {\doibase 10.1103/PhysRevLett.82.1963} {\bibfield  {journal}
  {\bibinfo  {journal} {Phys. Rev. Lett.}\ }\textbf {\bibinfo {volume} {82}},\
  \bibinfo {pages} {1963} (\bibinfo {year} {1999})}\BibitemShut {NoStop}%
\bibitem [{\citenamefont {Thornburg}\ \emph {et~al.}(1997)\citenamefont
  {Thornburg}, \citenamefont {M\"oller}, \citenamefont {Roy}, \citenamefont
  {Carr}, \citenamefont {Li},\ and\ \citenamefont {Erneux}}]{Thornburg1997}%
  \BibitemOpen
  \bibfield  {author} {\bibinfo {author} {\bibfnamefont {K.~S.}\ \bibnamefont
  {Thornburg}}, \bibinfo {author} {\bibfnamefont {M.}~\bibnamefont {M\"oller}},
  \bibinfo {author} {\bibfnamefont {R.}~\bibnamefont {Roy}}, \bibinfo {author}
  {\bibfnamefont {T.~W.}\ \bibnamefont {Carr}}, \bibinfo {author}
  {\bibfnamefont {R.-D.}\ \bibnamefont {Li}}, \ and\ \bibinfo {author}
  {\bibfnamefont {T.}~\bibnamefont {Erneux}},\ }\href {\doibase
  10.1103/PhysRevE.55.3865} {\bibfield  {journal} {\bibinfo  {journal} {Phys.
  Rev. E}\ }\textbf {\bibinfo {volume} {55}},\ \bibinfo {pages} {3865}
  (\bibinfo {year} {1997})}\BibitemShut {NoStop}%
\bibitem [{\citenamefont {Hohl}\ \emph {et~al.}(1999)\citenamefont {Hohl},
  \citenamefont {Gavrielides}, \citenamefont {Erneux},\ and\ \citenamefont
  {Kovanis}}]{Hohl1999}%
  \BibitemOpen
  \bibfield  {author} {\bibinfo {author} {\bibfnamefont {A.}~\bibnamefont
  {Hohl}}, \bibinfo {author} {\bibfnamefont {A.}~\bibnamefont {Gavrielides}},
  \bibinfo {author} {\bibfnamefont {T.}~\bibnamefont {Erneux}}, \ and\ \bibinfo
  {author} {\bibfnamefont {V.}~\bibnamefont {Kovanis}},\ }\href {\doibase
  10.1103/PhysRevA.59.3941} {\bibfield  {journal} {\bibinfo  {journal} {Phys.
  Rev. A}\ }\textbf {\bibinfo {volume} {59}},\ \bibinfo {pages} {3941}
  (\bibinfo {year} {1999})}\BibitemShut {NoStop}%
\bibitem [{\citenamefont {Kozyreff}\ \emph {et~al.}(2000)\citenamefont
  {Kozyreff}, \citenamefont {Vladimirov},\ and\ \citenamefont
  {Mandel}}]{Kozyreff2000}%
  \BibitemOpen
  \bibfield  {author} {\bibinfo {author} {\bibfnamefont {G.}~\bibnamefont
  {Kozyreff}}, \bibinfo {author} {\bibfnamefont {A.~G.}\ \bibnamefont
  {Vladimirov}}, \ and\ \bibinfo {author} {\bibfnamefont {P.}~\bibnamefont
  {Mandel}},\ }\href {\doibase 10.1103/PhysRevLett.85.3809} {\bibfield
  {journal} {\bibinfo  {journal} {Phys. Rev. Lett.}\ }\textbf {\bibinfo
  {volume} {85}},\ \bibinfo {pages} {3809} (\bibinfo {year}
  {2000})}\BibitemShut {NoStop}%
\bibitem [{\citenamefont {Allaria}\ \emph {et~al.}(2001)\citenamefont
  {Allaria}, \citenamefont {Arecchi}, \citenamefont {Di~Garbo},\ and\
  \citenamefont {Meucci}}]{Allaria2001}%
  \BibitemOpen
  \bibfield  {author} {\bibinfo {author} {\bibfnamefont {E.}~\bibnamefont
  {Allaria}}, \bibinfo {author} {\bibfnamefont {F.~T.}\ \bibnamefont
  {Arecchi}}, \bibinfo {author} {\bibfnamefont {A.}~\bibnamefont {Di~Garbo}}, \
  and\ \bibinfo {author} {\bibfnamefont {R.}~\bibnamefont {Meucci}},\ }\href
  {\doibase 10.1103/PhysRevLett.86.791} {\bibfield  {journal} {\bibinfo
  {journal} {Phys. Rev. Lett.}\ }\textbf {\bibinfo {volume} {86}},\ \bibinfo
  {pages} {791} (\bibinfo {year} {2001})}\BibitemShut {NoStop}%
\bibitem [{\citenamefont {Baas}\ \emph {et~al.}(2008)\citenamefont {Baas},
  \citenamefont {Lagoudakis}, \citenamefont {Richard}, \citenamefont {Andr\'e},
  \citenamefont {Dang},\ and\ \citenamefont {Deveaud-Pl\'edran}}]{Baas2008}%
  \BibitemOpen
  \bibfield  {author} {\bibinfo {author} {\bibfnamefont {A.}~\bibnamefont
  {Baas}}, \bibinfo {author} {\bibfnamefont {K.~G.}\ \bibnamefont
  {Lagoudakis}}, \bibinfo {author} {\bibfnamefont {M.}~\bibnamefont {Richard}},
  \bibinfo {author} {\bibfnamefont {R.}~\bibnamefont {Andr\'e}}, \bibinfo
  {author} {\bibfnamefont {L.~S.}\ \bibnamefont {Dang}}, \ and\ \bibinfo
  {author} {\bibfnamefont {B.}~\bibnamefont {Deveaud-Pl\'edran}},\ }\href
  {\doibase 10.1103/PhysRevLett.100.170401} {\bibfield  {journal} {\bibinfo
  {journal} {Phys. Rev. Lett.}\ }\textbf {\bibinfo {volume} {100}},\ \bibinfo
  {pages} {170401} (\bibinfo {year} {2008})}\BibitemShut {NoStop}%
\bibitem [{\citenamefont {Zhang}\ \emph {et~al.}(2012)\citenamefont {Zhang},
  \citenamefont {Wiederhecker}, \citenamefont {Manipatruni}, \citenamefont
  {Barnard}, \citenamefont {McEuen},\ and\ \citenamefont {Lipson}}]{Zhang2012}%
  \BibitemOpen
  \bibfield  {author} {\bibinfo {author} {\bibfnamefont {M.}~\bibnamefont
  {Zhang}}, \bibinfo {author} {\bibfnamefont {G.~S.}\ \bibnamefont
  {Wiederhecker}}, \bibinfo {author} {\bibfnamefont {S.}~\bibnamefont
  {Manipatruni}}, \bibinfo {author} {\bibfnamefont {A.}~\bibnamefont
  {Barnard}}, \bibinfo {author} {\bibfnamefont {P.}~\bibnamefont {McEuen}}, \
  and\ \bibinfo {author} {\bibfnamefont {M.}~\bibnamefont {Lipson}},\ }\href
  {\doibase 10.1103/PhysRevLett.109.233906} {\bibfield  {journal} {\bibinfo
  {journal} {Phys. Rev. Lett.}\ }\textbf {\bibinfo {volume} {109}},\ \bibinfo
  {pages} {233906} (\bibinfo {year} {2012})}\BibitemShut {NoStop}%
\bibitem [{\citenamefont {Bagheri}\ \emph {et~al.}(2013)\citenamefont
  {Bagheri}, \citenamefont {Poot}, \citenamefont {Fan}, \citenamefont
  {Marquardt},\ and\ \citenamefont {Tang}}]{Bagheri2013}%
  \BibitemOpen
  \bibfield  {author} {\bibinfo {author} {\bibfnamefont {M.}~\bibnamefont
  {Bagheri}}, \bibinfo {author} {\bibfnamefont {M.}~\bibnamefont {Poot}},
  \bibinfo {author} {\bibfnamefont {L.}~\bibnamefont {Fan}}, \bibinfo {author}
  {\bibfnamefont {F.}~\bibnamefont {Marquardt}}, \ and\ \bibinfo {author}
  {\bibfnamefont {H.~X.}\ \bibnamefont {Tang}},\ }\href {\doibase
  10.1103/PhysRevLett.111.213902} {\bibfield  {journal} {\bibinfo  {journal}
  {Phys. Rev. Lett.}\ }\textbf {\bibinfo {volume} {111}},\ \bibinfo {pages}
  {213902} (\bibinfo {year} {2013})}\BibitemShut {NoStop}%
\bibitem [{\citenamefont {Lee}\ and\ \citenamefont
  {Sadeghpour}(2013)}]{Lee2013}%
  \BibitemOpen
  \bibfield  {author} {\bibinfo {author} {\bibfnamefont {T.~E.}\ \bibnamefont
  {Lee}}\ and\ \bibinfo {author} {\bibfnamefont {H.~R.}\ \bibnamefont
  {Sadeghpour}},\ }\href {\doibase 10.1103/PhysRevLett.111.234101} {\bibfield
  {journal} {\bibinfo  {journal} {Phys. Rev. Lett.}\ }\textbf {\bibinfo
  {volume} {111}},\ \bibinfo {pages} {234101} (\bibinfo {year}
  {2013})}\BibitemShut {NoStop}%
\bibitem [{\citenamefont {Walter}\ \emph {et~al.}(2014)\citenamefont {Walter},
  \citenamefont {Nunnenkamp},\ and\ \citenamefont {Bruder}}]{Walter2014}%
  \BibitemOpen
  \bibfield  {author} {\bibinfo {author} {\bibfnamefont {S.}~\bibnamefont
  {Walter}}, \bibinfo {author} {\bibfnamefont {A.}~\bibnamefont {Nunnenkamp}},
  \ and\ \bibinfo {author} {\bibfnamefont {C.}~\bibnamefont {Bruder}},\ }\href
  {\doibase 10.1103/PhysRevLett.112.094102} {\bibfield  {journal} {\bibinfo
  {journal} {Phys. Rev. Lett.}\ }\textbf {\bibinfo {volume} {112}},\ \bibinfo
  {pages} {094102} (\bibinfo {year} {2014})}\BibitemShut {NoStop}%
\bibitem [{\citenamefont {Ohadi}\ \emph {et~al.}(2016)\citenamefont {Ohadi},
  \citenamefont {Gregory}, \citenamefont {Freegarde}, \citenamefont {Rubo},
  \citenamefont {Kavokin}, \citenamefont {Berloff},\ and\ \citenamefont
  {Lagoudakis}}]{Ohadi2016}%
  \BibitemOpen
  \bibfield  {author} {\bibinfo {author} {\bibfnamefont {H.}~\bibnamefont
  {Ohadi}}, \bibinfo {author} {\bibfnamefont {R.~L.}\ \bibnamefont {Gregory}},
  \bibinfo {author} {\bibfnamefont {T.}~\bibnamefont {Freegarde}}, \bibinfo
  {author} {\bibfnamefont {Y.~G.}\ \bibnamefont {Rubo}}, \bibinfo {author}
  {\bibfnamefont {A.~V.}\ \bibnamefont {Kavokin}}, \bibinfo {author}
  {\bibfnamefont {N.~G.}\ \bibnamefont {Berloff}}, \ and\ \bibinfo {author}
  {\bibfnamefont {P.~G.}\ \bibnamefont {Lagoudakis}},\ }\href {\doibase
  10.1103/PhysRevX.6.031032} {\bibfield  {journal} {\bibinfo  {journal} {Phys.
  Rev. X}\ }\textbf {\bibinfo {volume} {6}},\ \bibinfo {pages} {031032}
  (\bibinfo {year} {2016})}\BibitemShut {NoStop}%
\bibitem [{\citenamefont {Hillbrand}\ \emph {et~al.}(2020)\citenamefont
  {Hillbrand}, \citenamefont {Auth}, \citenamefont {Piccardo}, \citenamefont
  {Opa\ifmmode~\check{c}\else \v{c}\fi{}ak}, \citenamefont {Gornik},
  \citenamefont {Strasser}, \citenamefont {Capasso}, \citenamefont {Breuer},\
  and\ \citenamefont {Schwarz}}]{Hillbrand2020}%
  \BibitemOpen
  \bibfield  {author} {\bibinfo {author} {\bibfnamefont {J.}~\bibnamefont
  {Hillbrand}}, \bibinfo {author} {\bibfnamefont {D.}~\bibnamefont {Auth}},
  \bibinfo {author} {\bibfnamefont {M.}~\bibnamefont {Piccardo}}, \bibinfo
  {author} {\bibfnamefont {N.}~\bibnamefont {Opa\ifmmode~\check{c}\else
  \v{c}\fi{}ak}}, \bibinfo {author} {\bibfnamefont {E.}~\bibnamefont {Gornik}},
  \bibinfo {author} {\bibfnamefont {G.}~\bibnamefont {Strasser}}, \bibinfo
  {author} {\bibfnamefont {F.}~\bibnamefont {Capasso}}, \bibinfo {author}
  {\bibfnamefont {S.}~\bibnamefont {Breuer}}, \ and\ \bibinfo {author}
  {\bibfnamefont {B.}~\bibnamefont {Schwarz}},\ }\href {\doibase
  10.1103/PhysRevLett.124.023901} {\bibfield  {journal} {\bibinfo  {journal}
  {Phys. Rev. Lett.}\ }\textbf {\bibinfo {volume} {124}},\ \bibinfo {pages}
  {023901} (\bibinfo {year} {2020})}\BibitemShut {NoStop}%
\bibitem [{\citenamefont {Takemura}\ \emph {et~al.}(2020)\citenamefont
  {Takemura}, \citenamefont {Takiguchi}, \citenamefont {Sumikura},
  \citenamefont {Kuramochi}, \citenamefont {Shinya},\ and\ \citenamefont
  {Notomi}}]{Takemura2020}%
  \BibitemOpen
  \bibfield  {author} {\bibinfo {author} {\bibfnamefont {N.}~\bibnamefont
  {Takemura}}, \bibinfo {author} {\bibfnamefont {M.}~\bibnamefont {Takiguchi}},
  \bibinfo {author} {\bibfnamefont {H.}~\bibnamefont {Sumikura}}, \bibinfo
  {author} {\bibfnamefont {E.}~\bibnamefont {Kuramochi}}, \bibinfo {author}
  {\bibfnamefont {A.}~\bibnamefont {Shinya}}, \ and\ \bibinfo {author}
  {\bibfnamefont {M.}~\bibnamefont {Notomi}},\ }\href {\doibase
  10.1103/PhysRevA.102.011501} {\bibfield  {journal} {\bibinfo  {journal}
  {Phys. Rev. A}\ }\textbf {\bibinfo {volume} {102}},\ \bibinfo {pages}
  {011501} (\bibinfo {year} {2020})}\BibitemShut {NoStop}%
\bibitem [{\citenamefont {Stankovski}\ \emph {et~al.}(2017)\citenamefont
  {Stankovski}, \citenamefont {Pereira}, \citenamefont {McClintock},\ and\
  \citenamefont {Stefanovska}}]{Stankovski2017}%
  \BibitemOpen
  \bibfield  {author} {\bibinfo {author} {\bibfnamefont {T.}~\bibnamefont
  {Stankovski}}, \bibinfo {author} {\bibfnamefont {T.}~\bibnamefont {Pereira}},
  \bibinfo {author} {\bibfnamefont {P.~V.~E.}\ \bibnamefont {McClintock}}, \
  and\ \bibinfo {author} {\bibfnamefont {A.}~\bibnamefont {Stefanovska}},\
  }\href {\doibase 10.1103/RevModPhys.89.045001} {\bibfield  {journal}
  {\bibinfo  {journal} {Rev. Mod. Phys.}\ }\textbf {\bibinfo {volume} {89}},\
  \bibinfo {pages} {045001} (\bibinfo {year} {2017})}\BibitemShut {NoStop}%
\bibitem [{\citenamefont {Barclay}\ \emph {et~al.}(2005)\citenamefont
  {Barclay}, \citenamefont {Srinivasan},\ and\ \citenamefont
  {Painter}}]{Barclay2005}%
  \BibitemOpen
  \bibfield  {author} {\bibinfo {author} {\bibfnamefont {P.~E.}\ \bibnamefont
  {Barclay}}, \bibinfo {author} {\bibfnamefont {K.}~\bibnamefont {Srinivasan}},
  \ and\ \bibinfo {author} {\bibfnamefont {O.}~\bibnamefont {Painter}},\ }\href
  {\doibase 10.1364/OPEX.13.000801} {\bibfield  {journal} {\bibinfo  {journal}
  {Opt. Express}\ }\textbf {\bibinfo {volume} {13}},\ \bibinfo {pages} {801}
  (\bibinfo {year} {2005})}\BibitemShut {NoStop}%
\bibitem [{\citenamefont {Uesugi}\ \emph {et~al.}(2006)\citenamefont {Uesugi},
  \citenamefont {Song}, \citenamefont {Asano},\ and\ \citenamefont
  {Noda}}]{Uesugi2006}%
  \BibitemOpen
  \bibfield  {author} {\bibinfo {author} {\bibfnamefont {T.}~\bibnamefont
  {Uesugi}}, \bibinfo {author} {\bibfnamefont {B.-S.}\ \bibnamefont {Song}},
  \bibinfo {author} {\bibfnamefont {T.}~\bibnamefont {Asano}}, \ and\ \bibinfo
  {author} {\bibfnamefont {S.}~\bibnamefont {Noda}},\ }\href {\doibase
  10.1364/OPEX.14.000377} {\bibfield  {journal} {\bibinfo  {journal} {Opt.
  Express}\ }\textbf {\bibinfo {volume} {14}},\ \bibinfo {pages} {377}
  (\bibinfo {year} {2006})}\BibitemShut {NoStop}%
\bibitem [{\citenamefont {Leuthold}\ \emph {et~al.}(2010)\citenamefont
  {Leuthold}, \citenamefont {Koos},\ and\ \citenamefont
  {Freude}}]{Leuthold2010}%
  \BibitemOpen
  \bibfield  {author} {\bibinfo {author} {\bibfnamefont {J.}~\bibnamefont
  {Leuthold}}, \bibinfo {author} {\bibfnamefont {C.}~\bibnamefont {Koos}}, \
  and\ \bibinfo {author} {\bibfnamefont {W.}~\bibnamefont {Freude}},\ }\href
  {https://doi.org/10.1038/nphoton.2010.185} {\bibfield  {journal} {\bibinfo
  {journal} {Nature Photonics}\ }\textbf {\bibinfo {volume} {4}},\ \bibinfo
  {pages} {535 EP } (\bibinfo {year} {2010})},\ \bibinfo {note} {review
  Article}\BibitemShut {NoStop}%
\bibitem [{\citenamefont {Tanabe}\ \emph {et~al.}(2005)\citenamefont {Tanabe},
  \citenamefont {Notomi}, \citenamefont {Mitsugi}, \citenamefont {Shinya},\
  and\ \citenamefont {Kuramochi}}]{Tanabe2005}%
  \BibitemOpen
  \bibfield  {author} {\bibinfo {author} {\bibfnamefont {T.}~\bibnamefont
  {Tanabe}}, \bibinfo {author} {\bibfnamefont {M.}~\bibnamefont {Notomi}},
  \bibinfo {author} {\bibfnamefont {S.}~\bibnamefont {Mitsugi}}, \bibinfo
  {author} {\bibfnamefont {A.}~\bibnamefont {Shinya}}, \ and\ \bibinfo {author}
  {\bibfnamefont {E.}~\bibnamefont {Kuramochi}},\ }\href {\doibase
  10.1364/OL.30.002575} {\bibfield  {journal} {\bibinfo  {journal} {Opt.
  Lett.}\ }\textbf {\bibinfo {volume} {30}},\ \bibinfo {pages} {2575} (\bibinfo
  {year} {2005})}\BibitemShut {NoStop}%
\bibitem [{\citenamefont {Notomi}\ \emph {et~al.}(2005)\citenamefont {Notomi},
  \citenamefont {Shinya}, \citenamefont {Mitsugi}, \citenamefont {Kira},
  \citenamefont {Kuramochi},\ and\ \citenamefont {Tanabe}}]{Notomi2005}%
  \BibitemOpen
  \bibfield  {author} {\bibinfo {author} {\bibfnamefont {M.}~\bibnamefont
  {Notomi}}, \bibinfo {author} {\bibfnamefont {A.}~\bibnamefont {Shinya}},
  \bibinfo {author} {\bibfnamefont {S.}~\bibnamefont {Mitsugi}}, \bibinfo
  {author} {\bibfnamefont {G.}~\bibnamefont {Kira}}, \bibinfo {author}
  {\bibfnamefont {E.}~\bibnamefont {Kuramochi}}, \ and\ \bibinfo {author}
  {\bibfnamefont {T.}~\bibnamefont {Tanabe}},\ }\href {\doibase
  10.1364/OPEX.13.002678} {\bibfield  {journal} {\bibinfo  {journal} {Opt.
  Express}\ }\textbf {\bibinfo {volume} {13}},\ \bibinfo {pages} {2678}
  (\bibinfo {year} {2005})}\BibitemShut {NoStop}%
\bibitem [{\citenamefont {Weidner}\ \emph {et~al.}(2007)\citenamefont
  {Weidner}, \citenamefont {Combrié}, \citenamefont {de~Rossi}, \citenamefont
  {Tran},\ and\ \citenamefont {Cassette}}]{Weidner2007}%
  \BibitemOpen
  \bibfield  {author} {\bibinfo {author} {\bibfnamefont {E.}~\bibnamefont
  {Weidner}}, \bibinfo {author} {\bibfnamefont {S.}~\bibnamefont {Combrié}},
  \bibinfo {author} {\bibfnamefont {A.}~\bibnamefont {de~Rossi}}, \bibinfo
  {author} {\bibfnamefont {N.-V.-Q.}\ \bibnamefont {Tran}}, \ and\ \bibinfo
  {author} {\bibfnamefont {S.}~\bibnamefont {Cassette}},\ }\href {\doibase
  10.1063/1.2712502} {\bibfield  {journal} {\bibinfo  {journal} {Applied
  Physics Letters}\ }\textbf {\bibinfo {volume} {90}},\ \bibinfo {pages}
  {101118} (\bibinfo {year} {2007})},\ \Eprint
  {http://arxiv.org/abs/https://doi.org/10.1063/1.2712502}
  {https://doi.org/10.1063/1.2712502} \BibitemShut {NoStop}%
\bibitem [{\citenamefont {Haret}\ \emph {et~al.}(2009)\citenamefont {Haret},
  \citenamefont {Tanabe}, \citenamefont {Kuramochi},\ and\ \citenamefont
  {Notomi}}]{Haret2009}%
  \BibitemOpen
  \bibfield  {author} {\bibinfo {author} {\bibfnamefont {L.-D.}\ \bibnamefont
  {Haret}}, \bibinfo {author} {\bibfnamefont {T.}~\bibnamefont {Tanabe}},
  \bibinfo {author} {\bibfnamefont {E.}~\bibnamefont {Kuramochi}}, \ and\
  \bibinfo {author} {\bibfnamefont {M.}~\bibnamefont {Notomi}},\ }\href
  {\doibase 10.1364/OE.17.021108} {\bibfield  {journal} {\bibinfo  {journal}
  {Opt. Express}\ }\textbf {\bibinfo {volume} {17}},\ \bibinfo {pages} {21108}
  (\bibinfo {year} {2009})}\BibitemShut {NoStop}%
\bibitem [{\citenamefont {de~Rossi}\ \emph {et~al.}(2009)\citenamefont
  {de~Rossi}, \citenamefont {Lauritano}, \citenamefont {Combri\'e},
  \citenamefont {Tran},\ and\ \citenamefont {Husko}}]{Rossi2009}%
  \BibitemOpen
  \bibfield  {author} {\bibinfo {author} {\bibfnamefont {A.}~\bibnamefont
  {de~Rossi}}, \bibinfo {author} {\bibfnamefont {M.}~\bibnamefont {Lauritano}},
  \bibinfo {author} {\bibfnamefont {S.}~\bibnamefont {Combri\'e}}, \bibinfo
  {author} {\bibfnamefont {Q.~V.}\ \bibnamefont {Tran}}, \ and\ \bibinfo
  {author} {\bibfnamefont {C.}~\bibnamefont {Husko}},\ }\href {\doibase
  10.1103/PhysRevA.79.043818} {\bibfield  {journal} {\bibinfo  {journal} {Phys.
  Rev. A}\ }\textbf {\bibinfo {volume} {79}},\ \bibinfo {pages} {043818}
  (\bibinfo {year} {2009})}\BibitemShut {NoStop}%
\bibitem [{\citenamefont {Cazier}\ \emph {et~al.}(2013)\citenamefont {Cazier},
  \citenamefont {Checoury}, \citenamefont {Haret},\ and\ \citenamefont
  {Boucaud}}]{Cazier2013}%
  \BibitemOpen
  \bibfield  {author} {\bibinfo {author} {\bibfnamefont {N.}~\bibnamefont
  {Cazier}}, \bibinfo {author} {\bibfnamefont {X.}~\bibnamefont {Checoury}},
  \bibinfo {author} {\bibfnamefont {L.-D.}\ \bibnamefont {Haret}}, \ and\
  \bibinfo {author} {\bibfnamefont {P.}~\bibnamefont {Boucaud}},\ }\href
  {\doibase 10.1364/OE.21.013626} {\bibfield  {journal} {\bibinfo  {journal}
  {Opt. Express}\ }\textbf {\bibinfo {volume} {21}},\ \bibinfo {pages} {13626}
  (\bibinfo {year} {2013})}\BibitemShut {NoStop}%
\bibitem [{\citenamefont {Yacomotti}\ \emph {et~al.}(2006)\citenamefont
  {Yacomotti}, \citenamefont {Monnier}, \citenamefont {Raineri}, \citenamefont
  {Bakir}, \citenamefont {Seassal}, \citenamefont {Raj},\ and\ \citenamefont
  {Levenson}}]{Yacomotti2006}%
  \BibitemOpen
  \bibfield  {author} {\bibinfo {author} {\bibfnamefont {A.~M.}\ \bibnamefont
  {Yacomotti}}, \bibinfo {author} {\bibfnamefont {P.}~\bibnamefont {Monnier}},
  \bibinfo {author} {\bibfnamefont {F.}~\bibnamefont {Raineri}}, \bibinfo
  {author} {\bibfnamefont {B.~B.}\ \bibnamefont {Bakir}}, \bibinfo {author}
  {\bibfnamefont {C.}~\bibnamefont {Seassal}}, \bibinfo {author} {\bibfnamefont
  {R.}~\bibnamefont {Raj}}, \ and\ \bibinfo {author} {\bibfnamefont {J.~A.}\
  \bibnamefont {Levenson}},\ }\href {\doibase 10.1103/PhysRevLett.97.143904}
  {\bibfield  {journal} {\bibinfo  {journal} {Phys. Rev. Lett.}\ }\textbf
  {\bibinfo {volume} {97}},\ \bibinfo {pages} {143904} (\bibinfo {year}
  {2006})}\BibitemShut {NoStop}%
\bibitem [{\citenamefont {Brunstein}\ \emph {et~al.}(2012)\citenamefont
  {Brunstein}, \citenamefont {Yacomotti}, \citenamefont {Sagnes}, \citenamefont
  {Raineri}, \citenamefont {Bigot},\ and\ \citenamefont
  {Levenson}}]{Brunstein2012}%
  \BibitemOpen
  \bibfield  {author} {\bibinfo {author} {\bibfnamefont {M.}~\bibnamefont
  {Brunstein}}, \bibinfo {author} {\bibfnamefont {A.~M.}\ \bibnamefont
  {Yacomotti}}, \bibinfo {author} {\bibfnamefont {I.}~\bibnamefont {Sagnes}},
  \bibinfo {author} {\bibfnamefont {F.}~\bibnamefont {Raineri}}, \bibinfo
  {author} {\bibfnamefont {L.}~\bibnamefont {Bigot}}, \ and\ \bibinfo {author}
  {\bibfnamefont {A.}~\bibnamefont {Levenson}},\ }\href {\doibase
  10.1103/PhysRevA.85.031803} {\bibfield  {journal} {\bibinfo  {journal} {Phys.
  Rev. A}\ }\textbf {\bibinfo {volume} {85}},\ \bibinfo {pages} {031803}
  (\bibinfo {year} {2012})}\BibitemShut {NoStop}%
\bibitem [{\citenamefont {Matsuda}\ \emph {et~al.}(2011)\citenamefont
  {Matsuda}, \citenamefont {Kato}, \citenamefont {ichi Harada}, \citenamefont
  {Takesue}, \citenamefont {Kuramochi}, \citenamefont {Taniyama},\ and\
  \citenamefont {Notomi}}]{Matsuda2011}%
  \BibitemOpen
  \bibfield  {author} {\bibinfo {author} {\bibfnamefont {N.}~\bibnamefont
  {Matsuda}}, \bibinfo {author} {\bibfnamefont {T.}~\bibnamefont {Kato}},
  \bibinfo {author} {\bibfnamefont {K.}~\bibnamefont {ichi Harada}}, \bibinfo
  {author} {\bibfnamefont {H.}~\bibnamefont {Takesue}}, \bibinfo {author}
  {\bibfnamefont {E.}~\bibnamefont {Kuramochi}}, \bibinfo {author}
  {\bibfnamefont {H.}~\bibnamefont {Taniyama}}, \ and\ \bibinfo {author}
  {\bibfnamefont {M.}~\bibnamefont {Notomi}},\ }\href {\doibase
  10.1364/OE.19.019861} {\bibfield  {journal} {\bibinfo  {journal} {Opt.
  Express}\ }\textbf {\bibinfo {volume} {19}},\ \bibinfo {pages} {19861}
  (\bibinfo {year} {2011})}\BibitemShut {NoStop}%
\bibitem [{\citenamefont {Yang}\ \emph {et~al.}(2009)\citenamefont {Yang},
  \citenamefont {Yu}, \citenamefont {Kwong},\ and\ \citenamefont
  {Wong}}]{Yang2009}%
  \BibitemOpen
  \bibfield  {author} {\bibinfo {author} {\bibfnamefont {X.}~\bibnamefont
  {Yang}}, \bibinfo {author} {\bibfnamefont {M.}~\bibnamefont {Yu}}, \bibinfo
  {author} {\bibfnamefont {D.-L.}\ \bibnamefont {Kwong}}, \ and\ \bibinfo
  {author} {\bibfnamefont {C.~W.}\ \bibnamefont {Wong}},\ }\href {\doibase
  10.1103/PhysRevLett.102.173902} {\bibfield  {journal} {\bibinfo  {journal}
  {Phys. Rev. Lett.}\ }\textbf {\bibinfo {volume} {102}},\ \bibinfo {pages}
  {173902} (\bibinfo {year} {2009})}\BibitemShut {NoStop}%
\bibitem [{\citenamefont {Nozaki}\ \emph {et~al.}(2013)\citenamefont {Nozaki},
  \citenamefont {Shinya}, \citenamefont {Matsuo}, \citenamefont {Sato},
  \citenamefont {Kuramochi},\ and\ \citenamefont {Notomi}}]{Nozaki2013}%
  \BibitemOpen
  \bibfield  {author} {\bibinfo {author} {\bibfnamefont {K.}~\bibnamefont
  {Nozaki}}, \bibinfo {author} {\bibfnamefont {A.}~\bibnamefont {Shinya}},
  \bibinfo {author} {\bibfnamefont {S.}~\bibnamefont {Matsuo}}, \bibinfo
  {author} {\bibfnamefont {T.}~\bibnamefont {Sato}}, \bibinfo {author}
  {\bibfnamefont {E.}~\bibnamefont {Kuramochi}}, \ and\ \bibinfo {author}
  {\bibfnamefont {M.}~\bibnamefont {Notomi}},\ }\href {\doibase
  10.1364/OE.21.011877} {\bibfield  {journal} {\bibinfo  {journal} {Opt.
  Express}\ }\textbf {\bibinfo {volume} {21}},\ \bibinfo {pages} {11877}
  (\bibinfo {year} {2013})}\BibitemShut {NoStop}%
\bibitem [{\citenamefont {Flayac}\ \emph {et~al.}(2015)\citenamefont {Flayac},
  \citenamefont {Gerace},\ and\ \citenamefont {Savona}}]{Flayac2015}%
  \BibitemOpen
  \bibfield  {author} {\bibinfo {author} {\bibfnamefont {H.}~\bibnamefont
  {Flayac}}, \bibinfo {author} {\bibfnamefont {D.}~\bibnamefont {Gerace}}, \
  and\ \bibinfo {author} {\bibfnamefont {V.}~\bibnamefont {Savona}},\ }\href
  {\doibase 10.1038/srep11223} {\bibfield  {journal} {\bibinfo  {journal}
  {Scientific Reports}\ }\textbf {\bibinfo {volume} {5}},\ \bibinfo {pages}
  {11223} (\bibinfo {year} {2015})}\BibitemShut {NoStop}%
\bibitem [{\citenamefont {Yacomotti}\ \emph {et~al.}(2013)\citenamefont
  {Yacomotti}, \citenamefont {Haddadi},\ and\ \citenamefont
  {Barbay}}]{Yacomotti2013}%
  \BibitemOpen
  \bibfield  {author} {\bibinfo {author} {\bibfnamefont {A.~M.}\ \bibnamefont
  {Yacomotti}}, \bibinfo {author} {\bibfnamefont {S.}~\bibnamefont {Haddadi}},
  \ and\ \bibinfo {author} {\bibfnamefont {S.}~\bibnamefont {Barbay}},\ }\href
  {\doibase 10.1103/PhysRevA.87.041804} {\bibfield  {journal} {\bibinfo
  {journal} {Phys. Rev. A}\ }\textbf {\bibinfo {volume} {87}},\ \bibinfo
  {pages} {041804} (\bibinfo {year} {2013})}\BibitemShut {NoStop}%
\bibitem [{\citenamefont {Yu}\ \emph {et~al.}(2017)\citenamefont {Yu},
  \citenamefont {Xue}, \citenamefont {Semenova}, \citenamefont {Yvind},\ and\
  \citenamefont {Mork}}]{Yu2017}%
  \BibitemOpen
  \bibfield  {author} {\bibinfo {author} {\bibfnamefont {Y.}~\bibnamefont
  {Yu}}, \bibinfo {author} {\bibfnamefont {W.}~\bibnamefont {Xue}}, \bibinfo
  {author} {\bibfnamefont {E.}~\bibnamefont {Semenova}}, \bibinfo {author}
  {\bibfnamefont {K.}~\bibnamefont {Yvind}}, \ and\ \bibinfo {author}
  {\bibfnamefont {J.}~\bibnamefont {Mork}},\ }\href {\doibase
  10.1038/nphoton.2016.248} {\bibfield  {journal} {\bibinfo  {journal} {Nature
  Photonics}\ }\textbf {\bibinfo {volume} {11}},\ \bibinfo {pages} {81}
  (\bibinfo {year} {2017})}\BibitemShut {NoStop}%
\bibitem [{\citenamefont {Marconi}\ \emph {et~al.}(2020)\citenamefont
  {Marconi}, \citenamefont {Raineri}, \citenamefont {Levenson}, \citenamefont
  {Yacomotti}, \citenamefont {Javaloyes}, \citenamefont {Pan}, \citenamefont
  {Amili},\ and\ \citenamefont {Fainman}}]{Marconi2020}%
  \BibitemOpen
  \bibfield  {author} {\bibinfo {author} {\bibfnamefont {M.}~\bibnamefont
  {Marconi}}, \bibinfo {author} {\bibfnamefont {F.}~\bibnamefont {Raineri}},
  \bibinfo {author} {\bibfnamefont {A.}~\bibnamefont {Levenson}}, \bibinfo
  {author} {\bibfnamefont {A.~M.}\ \bibnamefont {Yacomotti}}, \bibinfo {author}
  {\bibfnamefont {J.}~\bibnamefont {Javaloyes}}, \bibinfo {author}
  {\bibfnamefont {S.~H.}\ \bibnamefont {Pan}}, \bibinfo {author} {\bibfnamefont
  {A.~E.}\ \bibnamefont {Amili}}, \ and\ \bibinfo {author} {\bibfnamefont
  {Y.}~\bibnamefont {Fainman}},\ }\href {\doibase
  10.1103/PhysRevLett.124.213602} {\bibfield  {journal} {\bibinfo  {journal}
  {Phys. Rev. Lett.}\ }\textbf {\bibinfo {volume} {124}},\ \bibinfo {pages}
  {213602} (\bibinfo {year} {2020})}\BibitemShut {NoStop}%
\bibitem [{\citenamefont {Bregni}(2002)}]{Bregni2002}%
  \BibitemOpen
  \bibfield  {author} {\bibinfo {author} {\bibfnamefont {S.}~\bibnamefont
  {Bregni}},\ }\href@noop {} {\emph {\bibinfo {title} {Synchronization of
  digital telecommunications networks}}},\ Vol.~\bibinfo {volume} {27}\
  (\bibinfo  {publisher} {Wiley New York},\ \bibinfo {year} {2002})\BibitemShut
  {NoStop}%
\bibitem [{\citenamefont {Kuramochi}\ \emph {et~al.}(2014)\citenamefont
  {Kuramochi}, \citenamefont {Grossman}, \citenamefont {Nozaki}, \citenamefont
  {Takeda}, \citenamefont {Shinya}, \citenamefont {Taniyama},\ and\
  \citenamefont {Notomi}}]{Kuramochi2014}%
  \BibitemOpen
  \bibfield  {author} {\bibinfo {author} {\bibfnamefont {E.}~\bibnamefont
  {Kuramochi}}, \bibinfo {author} {\bibfnamefont {E.}~\bibnamefont {Grossman}},
  \bibinfo {author} {\bibfnamefont {K.}~\bibnamefont {Nozaki}}, \bibinfo
  {author} {\bibfnamefont {K.}~\bibnamefont {Takeda}}, \bibinfo {author}
  {\bibfnamefont {A.}~\bibnamefont {Shinya}}, \bibinfo {author} {\bibfnamefont
  {H.}~\bibnamefont {Taniyama}}, \ and\ \bibinfo {author} {\bibfnamefont
  {M.}~\bibnamefont {Notomi}},\ }\href {\doibase 10.1364/OL.39.005780}
  {\bibfield  {journal} {\bibinfo  {journal} {Opt. Lett.}\ }\textbf {\bibinfo
  {volume} {39}},\ \bibinfo {pages} {5780} (\bibinfo {year}
  {2014})}\BibitemShut {NoStop}%
\bibitem [{\citenamefont {Van~Vaerenbergh}\ \emph {et~al.}(2012)\citenamefont
  {Van~Vaerenbergh}, \citenamefont {Fiers}, \citenamefont {Dambre},\ and\
  \citenamefont {Bienstman}}]{VanVaerenbergh2012}%
  \BibitemOpen
  \bibfield  {author} {\bibinfo {author} {\bibfnamefont {T.}~\bibnamefont
  {Van~Vaerenbergh}}, \bibinfo {author} {\bibfnamefont {M.}~\bibnamefont
  {Fiers}}, \bibinfo {author} {\bibfnamefont {J.}~\bibnamefont {Dambre}}, \
  and\ \bibinfo {author} {\bibfnamefont {P.}~\bibnamefont {Bienstman}},\ }\href
  {\doibase 10.1103/PhysRevA.86.063808} {\bibfield  {journal} {\bibinfo
  {journal} {Phys. Rev. A}\ }\textbf {\bibinfo {volume} {86}},\ \bibinfo
  {pages} {063808} (\bibinfo {year} {2012})}\BibitemShut {NoStop}%
\bibitem [{\citenamefont {Zhang}\ \emph {et~al.}(2013)\citenamefont {Zhang},
  \citenamefont {Fei}, \citenamefont {Cao}, \citenamefont {Cao}, \citenamefont
  {Xu},\ and\ \citenamefont {Chen}}]{Zhang2013}%
  \BibitemOpen
  \bibfield  {author} {\bibinfo {author} {\bibfnamefont {L.}~\bibnamefont
  {Zhang}}, \bibinfo {author} {\bibfnamefont {Y.}~\bibnamefont {Fei}}, \bibinfo
  {author} {\bibfnamefont {T.}~\bibnamefont {Cao}}, \bibinfo {author}
  {\bibfnamefont {Y.}~\bibnamefont {Cao}}, \bibinfo {author} {\bibfnamefont
  {Q.}~\bibnamefont {Xu}}, \ and\ \bibinfo {author} {\bibfnamefont
  {S.}~\bibnamefont {Chen}},\ }\href {\doibase 10.1103/PhysRevA.87.053805}
  {\bibfield  {journal} {\bibinfo  {journal} {Phys. Rev. A}\ }\textbf {\bibinfo
  {volume} {87}},\ \bibinfo {pages} {053805} (\bibinfo {year}
  {2013})}\BibitemShut {NoStop}%
\bibitem [{\citenamefont {Tanabe}\ \emph {et~al.}(2008)\citenamefont {Tanabe},
  \citenamefont {Taniyama},\ and\ \citenamefont {Notomi}}]{Tanabe2008}%
  \BibitemOpen
  \bibfield  {author} {\bibinfo {author} {\bibfnamefont {T.}~\bibnamefont
  {Tanabe}}, \bibinfo {author} {\bibfnamefont {H.}~\bibnamefont {Taniyama}}, \
  and\ \bibinfo {author} {\bibfnamefont {M.}~\bibnamefont {Notomi}},\ }\href
  {http://jlt.osa.org/abstract.cfm?URI=jlt-26-11-1396} {\bibfield  {journal}
  {\bibinfo  {journal} {J. Lightwave Technol.}\ }\textbf {\bibinfo {volume}
  {26}},\ \bibinfo {pages} {1396} (\bibinfo {year} {2008})}\BibitemShut
  {NoStop}%
\bibitem [{\citenamefont {Strogatz}(2018)}]{Strogatz2018}%
  \BibitemOpen
  \bibfield  {author} {\bibinfo {author} {\bibfnamefont {S.~H.}\ \bibnamefont
  {Strogatz}},\ }\href@noop {} {\emph {\bibinfo {title} {Nonlinear dynamics and
  chaos: with applications to physics, biology, chemistry, and engineering}}}\
  (\bibinfo  {publisher} {CRC Press},\ \bibinfo {year} {2018})\BibitemShut
  {NoStop}%
\bibitem [{\citenamefont {Nakao}(2016)}]{Nakao2016}%
  \BibitemOpen
  \bibfield  {author} {\bibinfo {author} {\bibfnamefont {H.}~\bibnamefont
  {Nakao}},\ }\href {\doibase 10.1080/00107514.2015.1094987} {\bibfield
  {journal} {\bibinfo  {journal} {Contemporary Physics}\ }\textbf {\bibinfo
  {volume} {57}},\ \bibinfo {pages} {188} (\bibinfo {year} {2016})},\ \Eprint
  {http://arxiv.org/abs/https://doi.org/10.1080/00107514.2015.1094987}
  {https://doi.org/10.1080/00107514.2015.1094987} \BibitemShut {NoStop}%
\bibitem [{\citenamefont {Ermentrout}(1996)}]{Ermentrout1996}%
  \BibitemOpen
  \bibfield  {author} {\bibinfo {author} {\bibfnamefont {B.}~\bibnamefont
  {Ermentrout}},\ }\href {\doibase 10.1162/neco.1996.8.5.979} {\bibfield
  {journal} {\bibinfo  {journal} {Neural Computation}\ }\textbf {\bibinfo
  {volume} {8}},\ \bibinfo {pages} {979} (\bibinfo {year} {1996})},\ \Eprint
  {http://arxiv.org/abs/https://doi.org/10.1162/neco.1996.8.5.979}
  {https://doi.org/10.1162/neco.1996.8.5.979} \BibitemShut {NoStop}%
\bibitem [{\citenamefont {Siegman}(1986)}]{Siegman1986}%
  \BibitemOpen
  \bibfield  {author} {\bibinfo {author} {\bibfnamefont {A.~E.}\ \bibnamefont
  {Siegman}},\ }\href@noop {} {\emph {\bibinfo {title} {Lasers}}}\ (\bibinfo
  {publisher} {University Science Mill Valley, Calif.},\ \bibinfo {year}
  {1986})\BibitemShut {NoStop}%
\bibitem [{\citenamefont {{Kurtz}}\ \emph {et~al.}(2005)\citenamefont
  {{Kurtz}}, \citenamefont {{Pradhan}}, \citenamefont {{Tun}}, \citenamefont
  {{Aye}}, \citenamefont {{Savant}}, \citenamefont {{Jannson}},\ and\
  \citenamefont {{DeShazer}}}]{Kurtz2005}%
  \BibitemOpen
  \bibfield  {author} {\bibinfo {author} {\bibfnamefont {R.~M.}\ \bibnamefont
  {{Kurtz}}}, \bibinfo {author} {\bibfnamefont {R.~D.}\ \bibnamefont
  {{Pradhan}}}, \bibinfo {author} {\bibfnamefont {N.}~\bibnamefont {{Tun}}},
  \bibinfo {author} {\bibfnamefont {T.~M.}\ \bibnamefont {{Aye}}}, \bibinfo
  {author} {\bibfnamefont {G.~D.}\ \bibnamefont {{Savant}}}, \bibinfo {author}
  {\bibfnamefont {T.~P.}\ \bibnamefont {{Jannson}}}, \ and\ \bibinfo {author}
  {\bibfnamefont {L.~G.}\ \bibnamefont {{DeShazer}}},\ }\href@noop {}
  {\bibfield  {journal} {\bibinfo  {journal} {IEEE Journal of Selected Topics
  in Quantum Electronics}\ }\textbf {\bibinfo {volume} {11}},\ \bibinfo {pages}
  {578} (\bibinfo {year} {2005})}\BibitemShut {NoStop}%
\bibitem [{\citenamefont {Notomi}\ \emph {et~al.}(2008)\citenamefont {Notomi},
  \citenamefont {Kuramochi},\ and\ \citenamefont {Tanabe}}]{Notomi2008}%
  \BibitemOpen
  \bibfield  {author} {\bibinfo {author} {\bibfnamefont {M.}~\bibnamefont
  {Notomi}}, \bibinfo {author} {\bibfnamefont {E.}~\bibnamefont {Kuramochi}}, \
  and\ \bibinfo {author} {\bibfnamefont {T.}~\bibnamefont {Tanabe}},\ }\href
  {\doibase 10.1038/nphoton.2008.226} {\bibfield  {journal} {\bibinfo
  {journal} {Nature Photonics}\ }\textbf {\bibinfo {volume} {2}},\ \bibinfo
  {pages} {741} (\bibinfo {year} {2008})}\BibitemShut {NoStop}%
\bibitem [{\citenamefont {Haddadi}\ \emph {et~al.}(2014)\citenamefont
  {Haddadi}, \citenamefont {Hamel}, \citenamefont {Beaudoin}, \citenamefont
  {Sagnes}, \citenamefont {Sauvan}, \citenamefont {Lalanne}, \citenamefont
  {Levenson},\ and\ \citenamefont {Yacomotti}}]{Haddadi2014}%
  \BibitemOpen
  \bibfield  {author} {\bibinfo {author} {\bibfnamefont {S.}~\bibnamefont
  {Haddadi}}, \bibinfo {author} {\bibfnamefont {P.}~\bibnamefont {Hamel}},
  \bibinfo {author} {\bibfnamefont {G.}~\bibnamefont {Beaudoin}}, \bibinfo
  {author} {\bibfnamefont {I.}~\bibnamefont {Sagnes}}, \bibinfo {author}
  {\bibfnamefont {C.}~\bibnamefont {Sauvan}}, \bibinfo {author} {\bibfnamefont
  {P.}~\bibnamefont {Lalanne}}, \bibinfo {author} {\bibfnamefont {J.~A.}\
  \bibnamefont {Levenson}}, \ and\ \bibinfo {author} {\bibfnamefont {A.~M.}\
  \bibnamefont {Yacomotti}},\ }\href {\doibase 10.1364/OE.22.012359} {\bibfield
   {journal} {\bibinfo  {journal} {Opt. Express}\ }\textbf {\bibinfo {volume}
  {22}},\ \bibinfo {pages} {12359} (\bibinfo {year} {2014})}\BibitemShut
  {NoStop}%
\bibitem [{\citenamefont {Priem}\ \emph {et~al.}(2005)\citenamefont {Priem},
  \citenamefont {Dumon}, \citenamefont {Bogaerts}, \citenamefont {Thourhout},
  \citenamefont {Morthier},\ and\ \citenamefont {Baets}}]{Priem2005}%
  \BibitemOpen
  \bibfield  {author} {\bibinfo {author} {\bibfnamefont {G.}~\bibnamefont
  {Priem}}, \bibinfo {author} {\bibfnamefont {P.}~\bibnamefont {Dumon}},
  \bibinfo {author} {\bibfnamefont {W.}~\bibnamefont {Bogaerts}}, \bibinfo
  {author} {\bibfnamefont {D.~V.}\ \bibnamefont {Thourhout}}, \bibinfo {author}
  {\bibfnamefont {G.}~\bibnamefont {Morthier}}, \ and\ \bibinfo {author}
  {\bibfnamefont {R.}~\bibnamefont {Baets}},\ }\href {\doibase
  10.1364/OPEX.13.009623} {\bibfield  {journal} {\bibinfo  {journal} {Opt.
  Express}\ }\textbf {\bibinfo {volume} {13}},\ \bibinfo {pages} {9623}
  (\bibinfo {year} {2005})}\BibitemShut {NoStop}%
\bibitem [{\citenamefont {Johnson}\ \emph {et~al.}(2006)\citenamefont
  {Johnson}, \citenamefont {Borselli},\ and\ \citenamefont
  {Painter}}]{Johnson2006}%
  \BibitemOpen
  \bibfield  {author} {\bibinfo {author} {\bibfnamefont {T.~J.}\ \bibnamefont
  {Johnson}}, \bibinfo {author} {\bibfnamefont {M.}~\bibnamefont {Borselli}}, \
  and\ \bibinfo {author} {\bibfnamefont {O.}~\bibnamefont {Painter}},\ }\href
  {\doibase 10.1364/OPEX.14.000817} {\bibfield  {journal} {\bibinfo  {journal}
  {Opt. Express}\ }\textbf {\bibinfo {volume} {14}},\ \bibinfo {pages} {817}
  (\bibinfo {year} {2006})}\BibitemShut {NoStop}%
\bibitem [{\citenamefont {Pernice}\ \emph {et~al.}(2010)\citenamefont
  {Pernice}, \citenamefont {Li},\ and\ \citenamefont {Tang}}]{Pernice2010}%
  \BibitemOpen
  \bibfield  {author} {\bibinfo {author} {\bibfnamefont {W.~H.~P.}\
  \bibnamefont {Pernice}}, \bibinfo {author} {\bibfnamefont {M.}~\bibnamefont
  {Li}}, \ and\ \bibinfo {author} {\bibfnamefont {H.~X.}\ \bibnamefont
  {Tang}},\ }\href {\doibase 10.1364/OE.18.018438} {\bibfield  {journal}
  {\bibinfo  {journal} {Opt. Express}\ }\textbf {\bibinfo {volume} {18}},\
  \bibinfo {pages} {18438} (\bibinfo {year} {2010})}\BibitemShut {NoStop}%
\bibitem [{\citenamefont {Xu}\ \emph {et~al.}(2019)\citenamefont {Xu},
  \citenamefont {Han}, \citenamefont {Lu}, \citenamefont {Gong}, \citenamefont
  {Qiu}, \citenamefont {Chen},\ and\ \citenamefont {Xiao}}]{Xu2019}%
  \BibitemOpen
  \bibfield  {author} {\bibinfo {author} {\bibfnamefont {D.}~\bibnamefont
  {Xu}}, \bibinfo {author} {\bibfnamefont {Z.-Z.}\ \bibnamefont {Han}},
  \bibinfo {author} {\bibfnamefont {Y.-K.}\ \bibnamefont {Lu}}, \bibinfo
  {author} {\bibfnamefont {Q.}~\bibnamefont {Gong}}, \bibinfo {author}
  {\bibfnamefont {C.-W.}\ \bibnamefont {Qiu}}, \bibinfo {author} {\bibfnamefont
  {G.}~\bibnamefont {Chen}}, \ and\ \bibinfo {author} {\bibfnamefont {Y.-F.}\
  \bibnamefont {Xiao}},\ }\href {\doibase 10.1117/1.AP.1.4.046002} {\bibfield
  {journal} {\bibinfo  {journal} {Advanced Photonics}\ }\textbf {\bibinfo
  {volume} {1}},\ \bibinfo {pages} {1 } (\bibinfo {year} {2019})}\BibitemShut
  {NoStop}%
\bibitem [{\citenamefont {Zheng}\ \emph {et~al.}(1998)\citenamefont {Zheng},
  \citenamefont {Hu},\ and\ \citenamefont {Hu}}]{Zheng1998}%
  \BibitemOpen
  \bibfield  {author} {\bibinfo {author} {\bibfnamefont {Z.}~\bibnamefont
  {Zheng}}, \bibinfo {author} {\bibfnamefont {G.}~\bibnamefont {Hu}}, \ and\
  \bibinfo {author} {\bibfnamefont {B.}~\bibnamefont {Hu}},\ }\href {\doibase
  10.1103/PhysRevLett.81.5318} {\bibfield  {journal} {\bibinfo  {journal}
  {Phys. Rev. Lett.}\ }\textbf {\bibinfo {volume} {81}},\ \bibinfo {pages}
  {5318} (\bibinfo {year} {1998})}\BibitemShut {NoStop}%
\bibitem [{\citenamefont {Daido}(1988)}]{Daido1988}%
  \BibitemOpen
  \bibfield  {author} {\bibinfo {author} {\bibfnamefont {H.}~\bibnamefont
  {Daido}},\ }\href {\doibase 10.1103/PhysRevLett.61.231} {\bibfield  {journal}
  {\bibinfo  {journal} {Phys. Rev. Lett.}\ }\textbf {\bibinfo {volume} {61}},\
  \bibinfo {pages} {231} (\bibinfo {year} {1988})}\BibitemShut {NoStop}%
\bibitem [{\citenamefont {Duport}\ \emph {et~al.}(2012)\citenamefont {Duport},
  \citenamefont {Schneider}, \citenamefont {Smerieri}, \citenamefont
  {Haelterman},\ and\ \citenamefont {Massar}}]{Duport2012}%
  \BibitemOpen
  \bibfield  {author} {\bibinfo {author} {\bibfnamefont {F.}~\bibnamefont
  {Duport}}, \bibinfo {author} {\bibfnamefont {B.}~\bibnamefont {Schneider}},
  \bibinfo {author} {\bibfnamefont {A.}~\bibnamefont {Smerieri}}, \bibinfo
  {author} {\bibfnamefont {M.}~\bibnamefont {Haelterman}}, \ and\ \bibinfo
  {author} {\bibfnamefont {S.}~\bibnamefont {Massar}},\ }\href {\doibase
  10.1364/OE.20.022783} {\bibfield  {journal} {\bibinfo  {journal} {Opt.
  Express}\ }\textbf {\bibinfo {volume} {20}},\ \bibinfo {pages} {22783}
  (\bibinfo {year} {2012})}\BibitemShut {NoStop}%
\end{thebibliography}

%

\end{document}